\pdfoutput=1

\documentclass[11pt]{article}
\usepackage{jheppub}
\usepackage{mathrsfs}
\usepackage{array,multirow}
\usepackage{arydshln}
\usepackage{ifpdf}

\newcommand{\cB}{\mathcal{B}}

\newcommand{\cF}{\mathcal{F}}

\newcommand{\cO}{\mathcal{O}}

\def\IP{\mathbb{P}}

\def\Label#1{\label{#1}%
  \smash{\hbox to0pt{\raise1ex\hbox{\tiny[#1]}\hss}}}
\def\noLabels{\let\Label=\label}
\def\nobbibitem{\let\bbibitem=\bibitem}
 \def\noBibitem{\let\Bibitem=\bibitem}

\let\SSS=\scriptstyle
\let\sss=\scriptscriptstyle
\let\ttt=\textstyle

\let\Tw=\widetilde
\let\0=\mathring
\let\2=\underline
\def\9{\vphantom{)}}

\DeclareMathOperator{\End}{End}

\let\into=\hookrightarrow
\let\onto=\twoheadrightarrow
\def\define{\overset{\text{def}}=}

\newtheorem{conj}{Conjecture}[section]
\def\IC{\mathbb{C}}
\def\fF{\mathfrak{F}}
\def\sF{\mathscr{F}}
\def\cK{\mathcal{K}}
\def\cP{\mathcal{P}}
\def\IP{\mathbb{P}}
\def\fP{\mathfrak{P}}
\def\cQ{\mathcal{Q}}

\def\ZZ{\mathbb{Z}}
\let\g=\gamma
\let\d=\delta

\def\rd{{\rm d}}
\let\vd=\partial
\let\e=\epsilon
\let\ve=\varepsilon
\let\f=\phi
\let\vf=\varphi
\let\k=\kappa
\let\vk=\varkappa
\let\l=\lambda
\let\m=\mu

\let\q=\theta
\let\vq=\vartheta
\let\s=\sigma
\let\W=\Omega
\let\too=\xrightarrow
\def\crlap#1{\rlap{$\vcenter{\hbox{$\scriptstyle#1$}}$}}
\def\doo#1{\downarrow\relax\!\crlap{#1}}

\def\mdot(#1,#2)#3#4{\put(#1,#2){\rotatebox{#3}{\multiput(0,0)(1,0){#4}{.}}}}
\def\vC#1{\vcenter{\hbox{\hss#1\hss}}}

\let\MC=\multicolumn
\let\MR=\multirow

\def\pMod#1{\,\,(\text{mod}\,\,#1)}

\def\str#1{\makebox[0pt][l]{\kern-1pt\color{red}\rule[.25ex]{#1mm}{1.5pt}}}
\def\CW#1{\raisebox{-.65ex}{\rotatebox{90}{$#1$}}}
\def\WC#1{\raisebox{1.75ex}{\rotatebox{-90}{$#1$}}}
\def\BW#1{\reflectbox{$#1$}}
\def\Cx#1{\makebox[0pt][c]{#1}}
\def\Lx#1{\makebox[0pt][l]{#1}}
\def\Rx#1{\makebox[0pt][r]{#1}}
\def\piC#1{\begin{picture}(0,0)#1\end{picture}}
\def\MM#1{\begin{matrix}#1\end{matrix}}
\def\mM#1{\begin{smallmatrix}#1\end{smallmatrix}}
\def\iM#1{\raisebox{2pt}{\kern-1.5mm\tiny$\begin{array}{l}#1\end{array}$}}
\def\pM#1{\!\left(\begin{smallmatrix}#1\end{smallmatrix}\right)}
\def\1{\mkern-1mu|\mkern5mu}
\def\bM#1{\!\left[\begin{smallmatrix}#1\end{smallmatrix}\right]}
\def\BM#1{\!\left[\begin{matrix}#1\end{matrix}\right]}

\def\K#1#2#3#4{\left#1\begin{array}{#2}#3\end{array}\right#4}
\def\ssK#1#2#3#4{\text{\scriptsize$\K{#1}{#2}{#3}{#4}$}}
\newcommand{\blue}{\textcolor{blue}}

\title{On Calabi-Yau generalized complete intersections\\
       from Hirzebruch varieties and novel K3-fibrations}

\author[a]{Per Berglund}
\author[b]{and Tristan H{\"u}bsch}

\affiliation[a]{Department of Physics, University of New Hampshire,\\
  Durham, NH 03824, USA}
\affiliation[b]{Department of Physics and Astronomy, Howard University,\\
  Washington, DC 20059, USA}

\emailAdd{per.berglund@unh.edu}
\emailAdd{thubsch@howard.edu}

\abstract{We consider the construction of Calabi-Yau varieties recently generalized to where the defining equations may have negative degrees over some projective space factors in the embedding space~\cite{rgCICY1}. Within such ``generalized complete intersection'' Calabi-Yau (``gCICY'') three-folds, we find several sequences of distinct manifolds.
 These include both novel elliptic and K3-fibrations and involve Hirzebruch surfaces and their higher dimensional analogues. En route, we generalize the standard techniques of cohomology computation to these generalized complete intersection Calabi-Yau varieties.}

\allowdisplaybreaks
\begin{document}
\noLabels 
\nobbibitem 
\noBibitem 

\makeatletter
\let\old@fpheader\@fpheader
\makeatother

\maketitle

\newpage

\section{Introduction, results and synopsis}
\label{s:IRS}
Ever since the discovery~\cite{rCHSW} that compact Calabi-Yau 3-folds provide string vacua
with possibly realistic phenomenology, the systematic construction of such varieties and computation of their physically relevant numerical characteristics has grown from the initial attempts~\cite{rH-CY0,rGHCYCI,rCYCI1,rCOK} to the impressive catalogue of some half a billion or more examples~\cite{rKreSka00b,rtCYdB}.
 Besides providing an incredible haystack of models in which to search for one that can describe the vacuum of our own Universe, this collection also provides a ``laboratory'' in which to explore both mathematical and physical properties of string theory, as well as M- and F-theory, such as mirror symmetry~\cite{rMMYau1,rMMYau2,rMMYau3,rCK,rMS}.

Recently, a novel class of ``generalized complete intersection Calabi-Yau'' (``gCICY'') 3-folds was introduced~\cite{rgCICY1}, as solutions to systems of algebraic equations in products of projective spaces where the defining equations may have negative degrees over some of the projective spaces. The Laurent polynomials of these equations are ``tuned'' so that their poles avoid the common zero-locus of the system, and this considerably enlarges the original pool of complete intersection Calabi-Yau (CICY) varieties~\cite{rH-CY0,rGHCYCI,rCYCI1}.
 In fact, we find that the construction of gCICYs, many of which are K3-fibrations, provides for even more distinct Calabi-Yau manifolds than reported in Ref.~\cite{rgCICY1}.

 In particular, the exploratory collection and preliminary classification in \cite{rgCICY1}  lists several sequences of K3-fibrations\footnote{The sequences in question are labeled as ``Type~III'' in Tables~1--4 of~\cite{rgCICY1}; see also Eqs.\,(5.28)--(5.32) therein. Following Ref.~\cite{rBeast}, we write $X\in[A||\mathbb{D}]$ to signify that $X$ is a generic member of the deformation family of varieties embedded in the {\em\/embedding space\/} $A$ by means of degree-$\mathbb{D}$ holomorphic constraints.} such as
\begin{equation}
  X_m \in
  \K{[}{c||c|c}{
   \IP^4 & 1 & 4 \\
   \IP^1 & m & 2{-}m \\}{]}^{(2,86)}_{-168},
  \qquad m=0,1,2,3,\ldots;\qquad
  X_m\subset F_m \in
  \K{[}{c||c}{
   \IP^4 & 1 \\
   \IP^1 & m \\}{]}.
 \label{e:Seq}
\end{equation}
 This defines the intermediate 4-fold $F_m$ as a degree-$\pM{1\\m}$ hypersurface $p(x,y)=0$ embedded in $A:=\IP^4\,{\times}\,\IP^1$, with $p(x,y)$ a holomorphic section of $\cP\define\cO\pM{1\\m\\}$.
 Then, $X_m\subset F_m$ is a degree-$\pM{4\\\!2-m\!}$ hypersurface $q(x,y)=0$, with $q(x,y)$ a holomorphic section of $\cQ\define\cO\pM{4\\\!2-m\!\\}$.
 For $m>2$, $q(x,y)$ is a Laurent-polynomial over $\IP^4\,{\times}\,\IP^1$, but may be chosen (``tuned'') such that its poles avoid the zeros of $p(x,y)$~\cite{rgCICY1}, making $X_m=\{p(x,y)=0\}\cap\{q(x,y)=0\}\subset\IP^4\,{\times}\IP^1$  well-defined for every $m\geqslant0$.

For the configurations~\eqref{e:Seq} with $m\leqslant2$ and other configurations in Appendix~\ref{s:More} with non-negative degrees, the classical analysis has been shown~\cite{rRes} to relate directly to the BRST treatment of constraints in the (world-sheet) field theory of superstrings compactified on so-defined Calabi-Yau 3-folds, and is also well known to correspond to Landau-Ginzburg orbifolds~\cite{rLGO} and Witten's gauged linear $\s$-model (GLSM)~\cite{rPhases}. For $m\geqslant3$ however, the (na{\"\i}ve) superpotentials in these world-sheet field theories would include Laurent polynomials in the fields, the effect of which in the GLSM is not understood. For this reason, we focus herein on the classical geometry and its physics implications, and defer the quantum aspects of compactification on such generalized complete intersections to a subsequent effort.

In Section~\ref{s:gCIs}, we show that 
the Calabi-Yau 3-folds constructed in this and similar semi-infinite sequences are in fact distinct from each other, and in physically relevant ways: Although all members within a sequence have the same Hodge numbers and even the same $\dim H^1(X_m,\End \,T)$, the {\em\/classical\/} triple intersections and the Pontryagin (Chern) evaluations of $H^{1,1}(X_m)\,{\approx}\,H^2(X_m,\ZZ)\,{\approx}\,H_4(X_m,\ZZ)$ elements vary within each sequence. 
 However,
we find that this $m$-dependence of {\em\/classical\/} topology characteristics is periodic in such ``Type~III'' sequences of K3-fibrations: in~\eqref{e:Seq} they depend on $m\pMod4$. In fact, this periodicity in the topological data of the Calabi-Yau 3-folds $X_m$ in~\eqref{e:Seq} is inherited from the 4-fold $F_m$.
Analogous phenomena are shown below to exist also in lower dimensions, generalizing the well-known $[m\pMod2]$-diffeomorphism in Hirzebruch surfaces $\fF_m$.  
 A theorem by C.T.C.~Wall~\cite{rWall} then guarantees that the sequence~\eqref{e:Seq} contains four distinct diffeomorphism classes of Calabi-Yau 3-folds, of which $X_3$ is a novel construction; see Figure~\ref{f:PinWheel} for a partial roadmap.
\begin{figure}[htp]
 \begin{center}
  \begin{picture}(150,50)(-75,-15)
   \put(-75,28){$h^{1,1}=2$, $h^{2,1}=86$; $\dim H^1(X_m,\End T)=188$}
   \put(6,.5){\Cx{\footnotesize$[\IP^4||5]$}}
   \put(-20,.5){\Rx{\footnotesize$[\IP^5||2,4]$}}
   \put(26,25){\Lx{\footnotesize$[\IP^4_{(1,1,1,1,4)}||8]$}}
   \put( 17, 13){\Cx{$\K{[}{c||cc}{\IP^4&1&4\\\IP^1&0&2}{]}$}}
   \put( 15.5, 28){\Cx{\footnotesize$\K{[}{c||cc}{\IP^3&4\\\IP^1&2}{]}$}}
   \put( 17,-10){\Cx{$\K{[}{c||cc}{\IP^4&1&4\\\IP^1&1&1}{]}$}}
   \put(-17,-10){\Cx{$\K{[}{c||cc}{\IP^4&1&4\\\IP^1&2&0}{]}$}}
   \put(-17, 13){\Cx{$\K{[}{c||c|c}{\IP^4&1&~~4\\\IP^1&3&-1}{]}$}}
   \put( 42, 13){\Cx{$\K{[}{c||c|c}{\IP^4&1&~~4\\\IP^1&4&-2}{]}$}}
   \put( 42,-10){\Cx{$\K{[}{c||c|c}{\IP^4&1&~~4\\\IP^1&5&-3}{]}$}}
   \put(-42,-10){\Cx{$\K{[}{c||c|c}{\IP^4&1&~~4\\\IP^1&6&-4}{]}$}}
   \put(-42, 13){\Cx{$\K{[}{c||c|c}{\IP^4&1&~~4\\\IP^1&7&-5}{]}$}}
   \put( 29,-10.25){\Cx{$\approx$}}
   \put( 29, 12.75){\Cx{$\approx$}}
   \put( 58,-10.25){\Cx{$\approx\cdots$}}
   \put( 58, 12.75){\Cx{$\approx\cdots$}}
   \put(-29,-10.25){\Cx{$\approx$}}
   \put(-29, 12.75){\Cx{$\approx$}}
   \put(-58,-10.25){\Cx{$\cdots\approx$}}
   \put(-58, 12.75){\Cx{$\cdots\approx$}}
   \put(16.5,20.5){\Cx{\rotatebox{90}{$=$}}}
  \color[rgb]{.6,0,.7} 
   \put(8,-1.5){\Cx{\rotatebox{-45}{$\rightleftharpoons$}}}
   \put(-26,-1.5){\Cx{\rotatebox{-45}{$\rightleftharpoons$}}}
   \put(25.5,19.5){\Cx{\rotatebox{60}{$\rightleftharpoons$}}}
   \put(23.5,24.5){\Cx{$\rightleftharpoons$}}
  \color[rgb]{0,.6,.2} 
   \put(16,6){\rotatebox{-90}{$\boldsymbol{\longrightarrow}$}}
   \put(39,6){\rotatebox{-90}{$\boldsymbol{\longrightarrow}$}}
   \put(-2,-10){\rotatebox{180}{$\boldsymbol{\to}$}}
   \put(-2, 12){\rotatebox{0}{$\boldsymbol{\to}$}}
   \put(-19,-1){\rotatebox{90}{$\boldsymbol{\longrightarrow}$}}
   \put(-45,-1){\rotatebox{90}{$\boldsymbol{\longrightarrow}$}}
  \color{blue}
  \end{picture}\vspace*{-5mm}
 \end{center}
 \caption{The $m\,{\color[rgb]{0,.6,.2}\to}\,[m{+}1\pMod4]$ ``pinwheel'' network of various models related in this article; see Section~\ref{s:gCIs}. Here,
 ``$\approx$'' denotes diffeomorphism as per Wall's theorem~\cite{rWall}, while
 ``{\color[rgb]{.8,0,.7}$\leftrightharpoons$}'' denotes conifold transitions such as discussed in Refs.~\cite{rCGH1,rCGH2}.}
 \label{f:PinWheel}
\end{figure}
In particular, distinct configurations in this network have conifold transitions to distinct $h^{1,1}=1$ models, as indicated in Figure~\ref{f:PinWheel}.
 As we show below, this distinction is related to the fact that the Hirzebruch surface $\fF_1$ {\em\/can\/} be blown down to $\IP^2$, while $\fF_0=\IP^1\,{\times}\,\IP^1$ cannot.

Also,
we provide a homological algebra explanation and general prescription for the specially tuned rational (Laurent-polynomial) sections $q(x,y)$ for realizing well-defined generalized complete intersections such as~\eqref{e:Seq} for $m\geqslant3$. This fully justifies and reconstructs the results of the iterative method reported in Ref.~\cite{rgCICY1}.
The sequence~\eqref{e:Seq} involves the 4-folds
 $F_m\in\ssK{[}{c||c}{\IP^4 & 1 \\ \IP^1 & m \\}{]}$, while other of the ``Type~III'' sequences of Ref.~\cite{rgCICY1} involve its 3- and 2-dimensional analogues. Adopting the name from the well-known 2-dimensional case, we dub these ($m$-twisted) ``Hirzebruch $n$-folds,''
 
In Section~\ref{s:Fms}, we analyze the so-constructed Calabi-Yau 3-folds $X_m$ as K3-fibrations, elliptic fibrations, and even iteratively nested fibrations; several of these features have been noted in Ref.~\cite{rgCICY1}. We also identify Calabi-Yau gCICY configurations which support this periodicity, and provide a geometric interpretation of this periodicity.

Just like the Hirzebruch surfaces $\fF_m$, Hirzebruch $n$-folds are no longer Fano for $m\geqslant2$. This explains the absence of such constructions from previous efforts, and the addition of some novel Calabi-Yau 3-folds even to such comprehensive databases as the Kreuzer-Skarke catalogue~\cite{rKreSka00b}. For example, the anticanonical bundle $\cQ=\cO\pM{4\\\!2-m\!\\}$ of $F_m$ in \eqref{e:Seq} is no longer positive over $\IP^1$ for $m\geqslant2$, and fails to be ample for $m\geqslant3$. Nevertheless, sequences such as~\eqref{e:Seq} do contain smooth and often novel Calabi-Yau 3-folds.
 
Finally, we summarize our results and their implications in Section~\ref{s:Coda}, and comment on the {\em\/quantum\/} cohomology of such 3-folds as well as finding their analogues in toric constructions.
Technical details are deferred to the appendices: in particular, Appendices~\ref{s:HFm} and~\ref{s:gTESS} collect the technical details of all requisite cohomology computations for the sequence of 3-folds~\eqref{e:Seq}, while Appendix C contains some further interesting examples, some of which have doubly periodic topological data.

\section{A curiously periodic sequence}
\label{s:gCIs}
We first explore the generalized complete intersections in projective spaces~\eqref{e:Seq}, and begin with a few key properties of the 4-folds $F_m$. To this end, we use the classical methods of algebraic geometry to compute the required cohomology of $X_m\subset F_m\subset A=\IP^4\,{\times}\,\IP^1$ iteratively; technical details are deferred to the Appendices~\ref{s:HFm} and~\ref{s:gTESS}. First, the 4-fold $F_m$ is defined:
\begin{equation}
  F_m\subset A=\IP^4\,{\times}\,\IP^1:\quad
  p(x,y)=p_{a\,(i_1\cdots\,i_m)}\,x^a\,y^{i_1}{\cdots}\,y^{i_m}=0,
 \label{e:fDef}
\end{equation}
where $(x^0\,{:}\cdots{:}\,x^4)\in\IP^4$ and $(y^0\,{:}\,y^1)\in\IP^1$ are the usual homogeneous coordinates, respectively, and $p_{a\,(i_1\cdots\,i_m)}$ is the defining tensor of $F_m$. Holomorphic sections and forms on $F_m$ are obtained by restricting from those on $A$ by means of the Koszul resolution {\em\/monad\/}\footnote{Throughout, ``$\into$'' denotes injections (1--1 linear maps which annihilate no non-trivial domain element), while ``$\onto$'' denotes surjections (``onto'' linear maps which omit nothing in the codomain).}:
\begin{equation}
  \cO_{\!A}\pM{\!-1\\\!-m\\} \overset{p}\into \cO_{\!A} \overset{\rho_F}\onto \cO_{F_m},
 \label{e:KFm}
\end{equation}
stating that sheaf of holomorphic functions on $F_m$ may be identified with the sheaf of holomorphic functions on $A$, taken modulo $p(x,y)$-multiples of $\cO_{\!A}\pM{\!-1\\\!-m\\}$-valued functions on $A$.
 We also use that $\IP^n=\frac{U(n+1)}{U(1)\times U(n)}$, whereby Bott-Borel-Weil's theorem guarantees that bundles over $\IP^n$ furnish $U(1)\,{\times}\,U(n)$-representations, all the cohomology groups valued in those bundles furnish $U(n{+}1)$-representations, and the maps in~\eqref{e:KFm} and between the associated cohomology groups are therefore completely represented by linear algebra with ``direct image'' $U(n{+}1)$-tensors~\cite{rBE}. For example,~\eqref{e:fDef} defines the tensor representative $p_{a\,(i_1\cdots\,i_m)}$ of sections of $\cO_{\!A}\pM{1\\m\\}$. This computational framework~\cite{rBE,rBeast} is also closely related to the Atiyah-Bott-G{\aa}rding-Candelas residue formulae~\cite{rABG,rPC,rRes}, as well as the BRST treatment of constraints and gauge-equivalence classes~\cite{rRes}.

\subsection{Viability of $X_m$}
To verify that the anticanonical bundle $\cK_{F_m}^*=\cQ$ of the hypersurface~\eqref{e:fDef} does have global holomorphic sections with which to define the Calabi-Yau hypersurface $X_m$, we compute the cohomology groups $H^*(F_m,\cQ)$ and find that $\dim H^0(F_m,\cQ)\geqslant105$ for all $m\geqslant0$; see Appendix~\ref{s:HFm}.
With that many linearly independent holomorphic sections to use for the defining equation of $X_m\subset F_m$, we expect that generic members of~\eqref{e:Seq} are smooth for each $m\geqslant0$, but we are not aware of a suitable generalization of Bertini's theorem to guarantee this also for the $m>2$ cases.\footnote{Ref.~\cite{rgCICY1} discusses computer-aided case-by-case methods of analysis which could do so for any fixed $m$, and cite the ``Type~III'' configuration~\eqref{e:Seq} as containing smooth models.} The analogous construction of 2-tori as hypersurfaces in Hirzebruch surfaces certainly does provide smooth models for all $m\geqslant0$; see Appendix~\ref{s:dFm}.

For now we assume that the system of 105 sections~\eqref{e:mH*Q} and~\eqref{e:4dFmK*} does suffice to construct smooth models $X_m\subset F_m\subset\IP^4\,{\times}\,\IP^1$ for every $m\geqslant0$. In turn, the existence of anticanonical holomorphic sections for gCICY's is not a foregone conclusion: there do exist similar gCICY sequences that however terminate, see Appendix~\ref{s:noSeq}.

\subsection{Hodge numbers}
\label{s:hpq}
The Calabi-Yau 3-folds $X_m$ in~\eqref{e:Seq} are defined by intersecting the hypersurface~\eqref{e:fDef} with a second one, defined by the vanishing of:
\begin{subequations}
 \label{e:qm}
\begin{alignat}9
 q(x,y)
 &=q_{(abcd)(ij)}\,x^a{\cdots}x^d\,y^iy^j\quad\&\quad
  q_{(abcd)\,i}\,x^a{\cdots}x^d\,y^i,&\quad& m&=0,1;\\
 &=\Big(q_{(abcd)} +q_{(abcd)\,k}^{\,j}\,\frac{y^k}{y^j}\Big)\,x^a{\cdots}x^d,
  &\quad& m&=2;\\
 &=q^{(j_1{\cdots}j_{m-2})}_{(abcd)}
    \frac{x^a{\cdots}x^d}{g^{(j_1{\cdots}j_{m-2})}(y)},&\quad& m&\geqslant3.
 \label{e:q>2}
\end{alignat}
\end{subequations}
As shown in Appendix~\ref{s:-K}, the defining tensors for the Laurent polynomials are parametrized by auxiliary $\IP^4$-cubics $f^{\,\cdots}_{(abc)}\,x^ax^bx^c$: 
\begin{subequations}
\label{e:q=efp}
\begin{alignat}9
  q_{(abcd)\,k}^{\,j}
  &\define\ve_{\9}^{ij}f_{(abc}\,p_{d)(ik)},&\quad m&=2; \label{e:2q=efp}\\
  q^{(j_1{\cdots}j_{m-2})}_{(abcd)}
  &\define\ve_{\9}^{i(j_1}f^{j_2{\cdots}j_{m-2})j_{m-1}{\cdots}j_{2m-3}}_{(abc}\,
             p^{\9}_{d)(i\,j_{m-1}{\cdots}j_{2m-3})},&\quad m&\geqslant3,
  \label{e:3q=efp}\\
 \text{where}
 &\hphantom{\define}~~f^{(j_1{\cdots}j_{2m-4})}_{(abc}\,p^{\9}_{d)(j_{m-3}{\cdots}j_{2m-4})}=0,&\quad m&\geqslant4. \label{e:4f}
\end{alignat}
\end{subequations}
The $(m{-}1)$ degree-$(m{-}2)$ generic $\IP^1$-polynomials $g^{(j_1{\cdots}j_{m-2})}(y)$ in~\eqref{e:q>2} allow separating the poles of $q(x,y)$ to $(m{-}1)(m{-}2)$ distinct locations and minimally extend the ``direct image'' linear algebra methods~\cite{rBE,rBeast}.

Using the adjunction relation $T_{X_m} \into T_A|_{X_m} \overset{\rd q}\onto [\cP\,{\oplus}\,\cQ]_{X_m}$, we compute the cohomology groups $H^*(X_m,T)=H^*(X_m,\wedge^2T^*)$. Deferring the technical details of the computation to Appendix~\ref{s:gTESS}, we quote here that
\begin{equation}
  h^{1,2}=\dim H^1(X_m,T)=86
   \qquad\text{and}\qquad
  h^{1,1}=h^{2,2}=\dim H^2(X_m,T)=2,
 \label{e:H**Fm}
\end{equation}
uniformly for all $m\geqslant0$.
 The same techniques also compute $\dim H^1(X_m,\text{End}\,T)=188$.
 
In particular, the results~\eqref{e:H**Fm} computed in Appendix~\ref{s:T} also prove that $H^1(X_m,T^*)=H^{1,1}(X_m)$, the dual of $H^2(X_m,T)$ is generated by (the pullbacks of) the K\"ahler classes $J_1$ of $\IP^4$ and $J_2$ of $\IP^1$ for all $m\geqslant0$.
The standard computation of the Chern class then gives:
\begin{equation}
 \begin{aligned}
   c(X_m) &=\frac{(1{+}J_1)^5(1{+}J_2)^2}
                 {(1{+}J_1{+}mJ_2)(1{+}4J_1+(2{-}m)J_2)},\\
   &=1 +\big(6J_1^{\,2}+(8{-}3m)J_1J_2\big)
        +\big({-}20J_1^{\,3}-(32{+}15m)J_1^{\,2}J_2\big),
 \label{e:ChXm}
 \end{aligned}
\end{equation}
confirming that the Euler number is independent of $m$:
\begin{equation}
 \chi_{_E} =\int_{X_m}\mkern-12mu c_3
  =\int_A\big(J_1{+}mJ_2\big)\big(4J_1+(2{-}m)J_2\big)\,c_3 = -168.
\end{equation}

\subsection{Classical topology}
\label{s:Y+P}
Wall's theorem~\cite{rWall} guarantees that the diffeomorphism class of compact and orientable real 6-dimensio\-nal manifolds $X$ is determined by the Betti numbers $b_2$ and $b_3$, the cubic intersections (classical Yukawa couplings) and the (first Pontryagin class) $p_1$-evaluation of $H^2(X,\ZZ)\approx H_4(X,\ZZ)$ elements; see the full discussion below.
The standard relation $p_1=c_1^{~2}{-}2c_2$ simplifies for Calabi-Yau 3-folds to $p_1=-2c_2$, and we also have that $b_2=h^{1,1}$ and $b_3=2+2h^{2,1}$. 

As shown in Appendix~\ref{s:T}, $H^{1,1}(X_m)\approx H^2(X_m,\ZZ)$ is generated by (the pullbacks to $X_m$ of) the K\"ahler classes of $\IP^4$ and $\IP^1$, so the classical Yukawa couplings in $H^{1,1}(X_m)$ are the standard classical (topological) intersection numbers:
\begin{alignat}9
  [(aJ_1+bJ_2)^3]_{X_m}
  &=2a^3+3a^2(\2{4b{+}ma}),&\quad\text{i.e.,}\quad
  &\Bigg\{\begin{array}{r@{\,=\,}l}
            \k_{111}&2{+}3m,\\[-1pt]
            \k_{112}&4,\\[-1pt]
            \k_{122}&0=\k_{222}.
          \end{array}
 \label{e:k3Xm}
\intertext{Also,}
  C_2[aJ_1+bJ_2]
  &= 44a+6(\2{4b{+}ma}),&\quad\text{i.e.,}\quad
  &\Big\{\begin{array}{r@{\,=\,}l}
            C_2[J_1]&44+6m,\\
            C_2[J_2]&24.
          \end{array}
 \label{e:p1Xm}
\end{alignat}
For the sequence of Calabi-Yau 3-folds~\eqref{e:Seq}, $b_2=2$ and $b_3=174$ are $m$-independent. In turn, the topological invariants~\eqref{e:k3Xm} and the Chern evaluation~\eqref{e:p1Xm} do depend on $m$, proving that the sequence~\eqref{e:Seq} does contain topologically distinct Calabi-Yau 3-folds.

Given that the Hodge diamond and the Euler characteristic are independent of $m$, and the topological intersections~\eqref{e:k3Xm} and Chern evaluations~\eqref{e:p1Xm} depend on $bJ_2$ and $m$ only through the (underlined) hallmark combination $(4b{+}am)$, 
it follows that all topological invariants remain unchanged by transforming $(a,b,m)\to(a,b{-}ac,m{+}4c)$, which is the integral basis-change relation:
\begin{equation}
 \bM{J_1\\J_2\\}_m \overset{\approx}{\longleftrightarrow}
  \bM{1&-c\\[2pt]0&~~1\\[2pt]}\bM{J_1\\J_2\\}_{m+4c},\qquad
 c\in\ZZ,\quad \det\bM{1&-c\\[2pt]0&~~1\\[2pt]}=1.
 \label{e:mod4}
\end{equation}
That is, the topological invariants~\eqref{e:k3Xm} and~\eqref{e:p1Xm} of $X_m$ and of $X_{m+4c}$ for $c\in\ZZ$ differ only by a basis change, and Wall's theorem guarantees that $X_m$ is diffeomorphic to $X_{m+4c}$ for all $c\in\ZZ$. Through this $[m\pMod4]$-dependence of the topological data~\eqref{e:k3Xm} and~\eqref{e:p1Xm}, Wall's theorem guarantees that the sequence~\eqref{e:Seq} contains precisely four distinct diffeomorphism classes of Calabi-Yau 3-folds, counted by $m\pMod4$. It is this $[m\pMod4]$-periodicity that provides the ``pinwheel'' diagram in Figure~\ref{f:PinWheel} with the characteristic cyclicality. 

We close here with a remark on the use of Wall's theorem.
 ({\bf1})~As the zero set of ample and positive line bundles, all $m$-twisted Hirzebruch $n$-folds are directly subject to the Lefschetz hyperplane theorem: $H^r(\sF^{\sss(n)},\ZZ)=H^r(\IP^n\,{\times}\,\IP^1,\ZZ)$ for $r\neq n$, and has no torsion.
 ({\bf2})~For $r=n$, the independent computation of the Euler number and the use of the universal coefficient theorem~\cite{rMcC-SS} jointly insure that also $H^n(\sF^{\sss(n)}_m,\ZZ)=H^n(\IP^n\,{\times}\,\IP^1,\ZZ)$, and has no torsion.
 ({\bf3})~The torsion-free (co)homology of all $m$-twisted Hirzebruch $n$-folds exhibits the $[m\pMod{n}]$-periodicity in the (classical) ring structure of $H^*(\sF^{\sss(n)}_m,\ZZ)$; see Appendices~\ref{s:OtherFm} and~\ref{s:dFm}.
 ({\bf4})~The Calabi-Yau $(n{-}1)$-folds embedded as anticanonical hypersurfaces in the $m$-twisted Hirzebruch $n$-folds exhibit exactly the same $[m\pMod{n}]$-periodicity --- generalizing~\eqref{e:k3Xm}-\eqref{e:p1Xm}, which therefore cannot possibly be the consequence of any torsion element.
Finally, 
 ({\bf5})~since the 2nd Stiefel-Whitney class $w_2$ is a $\ZZ_2$ reduction of the 1st Chern class --- which vanishes for Calabi-Yau $(n{-}1)$-folds by definition --- the $w_2=0$  condition of Wall's theorems is also satisfied.

\section{Calabi-Yau 3-folds from Hirzebruch $n$-folds}
\label{s:Fms}
All the ``Type-III'' sequences of Ref.~\cite{rgCICY1} involve Hirzebruch $n$-folds: Just as our main example~\eqref{e:Seq} involves the Hirzebruch 4-fold
 $F_m\in\ssK{[}{c||c}{\IP^4 & 1 \\ \IP^1 & m \\}{]}$, the sequence~\cite{rgCICY1}
\begin{equation}
 X'_m\in
 \K{[}{c||c|c}{\IP^1 & 0 & 2 \\[1pt] \hdashline[1pt/1pt]\rule{0pt}{12pt}
               \IP^3 & 1 & 3 \\
               \IP^1 & m & 2{-}m \\}{]}^{(3,75)}_{-144}
 \quad
 \begin{array}{r@{\,=\,}l}
   [(aJ_1{+}bJ_2{+}cJ_3)^3]_{X'_m} &6ab^2+2b(3a{+}b)(\2{3c{+}bm}),\\[2mm]
   C_2[(aJ_1{+}bJ_2{+}cJ_3)] &24a{+}36b+2(\2{3c{+}bm})],
 \end{array}
 \label{e:3dF:Xm}
\end{equation}
involves $\cF_m\in\ssK[{c||c}{\IP^3&1\\\IP^1&m\\}]$ while the ``Type~III'' sequences~\cite{rgCICY1}
\begin{subequations}
 \label{e:2dF:Xm}
\begin{alignat}9
   X''_m&\in
   \K{[}{c||c|c}{\IP^2 & 0 & 3 \\[1pt] \hdashline[1pt/1pt]\rule{0pt}{12pt}
                 \IP^2 & 1 & 2 \\
                 \IP^1 & m & 2{-}m \\}{]}^{(3,75)}_{-144}
 &\quad&
 \begin{array}{r@{\,=\,}l}
   [(aJ_1{+}bJ_2{+}cJ_3)^3]_{X''_m} &6a^2b + 3a(a{+}3b)(\2{2c{+}bm}),\\[2mm]
   C_2[(aJ_1{+}bJ_2{+}cJ_3)] &36a{+}24b + 12(\2{2c{+}bm}),
 \end{array}
 \\
   X'''_m&\in
   \K{[}{c||c|c}{\IP^1 & 0 & 2 \\
                 \IP^1 & 0 & 2 \\[1pt] \hdashline[1pt/1pt]\rule{0pt}{12pt}
                 \IP^2 & 1 & 2 \\
                 \IP^1 & m & 2{-}m \\}{]}^{(4,68)}_{-128}
 &\quad&
 \begin{array}{r@{}l}
   [(aJ_1{+}bJ_2{+}&cJ_3+dJ_4)^3]_{X'''_m}\\
     &= 12abc + 6(ab{+}ac{+}bc)(\2{2d{+}cm}),\\[2mm]
   C_2[(aJ_1{+}bJ_2{+}&cJ_3+dJ_4)]\\
     &= 24(a{+}b{+}c) + 12(\2{2d{+}cm}),\\
 \end{array}
\end{alignat}
\end{subequations}
both involve $\fF_m\in\ssK{[}{c||c}{\IP^2 & 1 \\ \IP^1 & m \\}{]}$, which are well-known as Hirzebruch surfaces~\cite{rBeast,rGE-CCAG} as well as ``rational ruled surfaces''~\cite{rGrHa}. 
 
\subsection{Fibrations} 
Each $X'_m$ in the sequence~\eqref{e:3dF:Xm} is a generalized ``double solid''~\cite{rC-2x3}: the small resolution\footnote{The branching locus is itself typically singular even if the whole 3-fold is smooth~\cite[p.\,141]{rBeast}.} of a double-cover of the Hirzebruch 3-fold $\cF_m$, branched over a degree-$\pM{6\\\!4-2m\!\\}$ hypersurface $\cB\subset\cF_m$. Notice that the branching locus is itself a generalized complete intersection from the configuration $\ssK[{c||c|c}{\IP^3&1&6\\\IP^1&m&4{-}2m\\}]$, where the second constraint becomes negative over $\IP^1$ for $m\geqslant3$. It is amusing to think of~\eqref{e:3dF:Xm} also as a deformation family of ``see-saw twisted''\footnote{For $m\geqslant3$, the twist of the fibration is positive over $\IP^3$ but negative over $\IP^1$.} double-point fibrations over $\cF_m$, since $[\IP^1\|2]=\{\text{2 pts}\}$ is the Calabi-Yau 0-fold. In turn, we may also regard~\eqref{e:3dF:Xm} as a deformation family of fibrations of $\text{K3}\in\ssK[{c||c|c}{\IP^3 &1 &3\\ \IP^1 &m &2{-}m\\}]$ over $\IP^1$, where now fibers are K3 surfaces, the well-known Calabi-Yau 2-folds.

 In turn, the two sequences of Calabi-Yau 3-folds~\eqref{e:2dF:Xm} may be regarded as ``ordinary'' elliptic fibrations over the Fano (del Pezzo) bases $\IP^2$ and $\IP^1\,{\times}\,\IP^1$, respectively, however with the fibers being ``generalized complete intersection'' tori in $\ssK[{c||c|c}{\IP^2&1&2\\\IP^1&m&2{-}m\\}]$. In turn, the same sequences may also be regarded as (see-saw complementarily twisted for $m>2$) elliptic (torus) fibrations over the Hirzebruch surfaces $\fF_m$, where the fibers are familiar tori from the configurations $[\IP^2\|3]$ and $\ssK[{c||c}{\IP^1&2\\\IP^1&2\\}]$, respectively. Viewed this way and since the Hirzebruch surfaces $\fF_m$ are themselves fibrations, the sequences~\eqref{e:2dF:Xm} are in fact iterated fibrations. 

In agreement with Ref.~\cite{rgCICY1}, we find that each of these Calabi-Yau 3-folds may be regarded as a fibration of a Calabi-Yau $n$-folds over a $(3{-}n)$-dimensional base in at least two different ways. As compared with the constructions studied until Ref.~\cite{rgCICY1}, the novelty in~\eqref{e:Seq}, (\ref{e:3dF:Xm}) and~\eqref{e:2dF:Xm} stems either from:
 ({\small\bf1})~using a decidedly non-Fano base such as the 3-folds $\cF_m$ and the 2-folds $\fF_m$ for $m\geqslant3$, or from
 ({\small\bf2})~fibering generalized complete intersection Calabi-Yau $n$-folds such as
\begin{equation}
  T^2\in\K[{c||c|c}{\IP^2&1&2\\\IP^1&m&2{-}m\\}]\qquad\text{and}\qquad
  \text{K3}\in\K[{c||c|c}{\IP^3&1&3\\\IP^1&m&2{-}m\\}],\qquad
  \text{with}~~m\geqslant0.
\end{equation}

The novelty of these fibrations is seen already in our main example, the sequence~\eqref{e:Seq}, which contains four distinct diffeomorphism classes of Calabi-Yau 3-folds represented by the configurations
\begin{equation}
  \K[{c||c|c}{\IP^4&1&4\\\hdashline[1pt/1pt] \IP^1&0&2\\}]
   =\K[{c||c}{\IP^3&4\\\hdashline[1pt/1pt] \IP^1&2\\}],\qquad
  \K[{c||c|c}{\IP^4&1&4\\\hdashline[1pt/1pt] \IP^1&1&1\\}],\qquad
  \K[{c||c|c}{\IP^4&1&4\\\hdashline[1pt/1pt] \IP^1&2&0\\}],\qquad
  \K[{c||c|c}{\IP^4&1&~~4\\\hdashline[1pt/1pt] \IP^1&3&-1\\}].
\end{equation}
In the first of these, the first, degree-$\pM{1\\0\\}$ defining equation is a simple, $\IP^1$-constant hyperplane, $[\IP^4||1]\approx\IP^3$, which induces a global isomorphism indicated by the ``='' sign. The resulting configuration, $\ssK[{c||c}{\IP^3&4\\\IP^1&2}]$ may be regarded as a deformation family of quadratic $\text{K3}\in[\IP^3||4]$-fibrations over $\IP^2$.
 In turn, the remaining three representatives may also be regarded as K3-fibrations over $\IP^1$ but in subtly different ways --- which explains the $m$-dependence in the intersection numbers~\eqref{e:k3Xm} and Chern evaluations~\eqref{e:p1Xm}.
 
In particular, the second configuration is a fibration of a $\text{K3}\in[\IP^4||1,4]$ surface defined as the intersection of a hyperplane and a quartic in $\IP^4$ --- both of which vary non-trivially (linearly) over the base $\IP^1$. Although a hyperplane in $\IP^4$ is always isomorphic to a $\IP^3$, now both the hyperplane and the quartic vary over the base $\IP^1$. Thus, the isomorphism $[\IP^4||1]\approx\IP^3$ keeps also varying over the base $\IP^1$, so that this is not a fibration (over $\IP^1$) of a quartic hypersurface in a fixed $\IP^3$, but in a similarly $\IP^1$-variable $\IP^3\in[\IP^4||1]$.

The third configuration now has the quartic $[\IP^3||4]$ held constant over $\IP^1$, and is being intersected by a (quadratically) $\IP^1$-variable hyperplane in $\IP^4$.

Finally, the fourth configuration again has both the hyperplane and the quartic vary over the base $\IP^1$, but differently than in the second configuration: the hyperplane now varies cubically, while the quartic varies ``inverse-linearly'' (of degree-$(-1)$) over $\IP^1$.

Viewing the succession of these various types of fibration, the classical $[m\pMod4]$-periodicity~\eqref{e:mod4} is rather surprising. For example, the configurations
\begin{equation}
  \K[{c||c}{\IP^3&4\\\hdashline[1pt/1pt] \IP^1&2\\}]=
   \K[{c||c|c}{\IP^4&1&4\\\hdashline[1pt/1pt] \IP^1&0&2\\}],\quad
  \K[{c||c|c}{\IP^4&1&~~4\\\hdashline[1pt/1pt] \IP^1&4&-2\\}],\quad
  \K[{c||c|c}{\IP^4&1&~~4\\\hdashline[1pt/1pt] \IP^1&8&-6\\}],\quad
  \K[{c||c|c}{\IP^4&~1&~~~4\\\hdashline[1pt/1pt] \IP^1&12&-10\\}],\quad\ldots
 \label{e:m=0}
\end{equation}
are all deformation families of Calabi-Yau 3-folds that are diffeomorphic to each other by virtue of Wall's theorem, in spite of the increasingly higher degree of $\IP^1$-fibration of the hyperplane in $\IP^4$ complemented by the increasingly more negative degree of $\IP^1$-fibration of the quartic in $\IP^4$.

\subsection{Periodicity}
The peculiar $[m\pMod4]$-periodic diffeomorphisms $X_m\approx X_{m+4}$~\eqref{e:mod4} of the Calabi-Yau 3-folds in~\eqref{e:Seq} in fact stem from the same diffeomorphisms~\eqref{e:Fmod4} between the Hirzebruch 4-folds $F_m$; see Appendix~\ref{s:HFm} for more detail. Indeed, the $[m\pMod{n}]$-periodic diffeomorphisms of the Hirzebruch $n$-folds induce the same periodicity in all ``Type-III'' sequences:
\begin{alignat}9
 \text{$[m\pMod4]$:}&\quad
 F_m
 &\in\K[{c||c}{\IP^4&1\\ \IP^1&m\\}]
 &~~\text{and}&\quad
 X_m
 &\in\K[{c||c|c}{\IP^4 & 1 & \piC{\put(1,1){\color{blue}\circle{5}}}4\\
                 \IP^1&m&2{-}m\\}];\\
 \text{$[m\pMod3]$:}&\quad
 \cF_m
 &\in\K[{c||c}{\IP^3&1\\ \IP^1&m\\}]
 &~~\text{and}&\quad
 X'_m
 &\in\K[{c||c|c}{\IP^1 & 0 & 2 \\[1pt]
                  \hdashline[1pt/1pt]\rule{0pt}{12pt}
                 \IP^3 & 1 & \piC{\put(1,1){\color{blue}\circle{5}}}3 \\
                 \IP^1 & m & 2{-}m \\}];\\
 \text{$[m\pMod2]$:}&\quad
 \fF_m
 &\in\K[{c||c}{\IP^2&1\\ \IP^1&m\\}]
 &~~\text{and}&\quad
 X''_m
 &\in\K[{c||c|c}{\IP^2 & 0 & 3 \\[1pt]
                  \hdashline[1pt/1pt]\rule{0pt}{12pt}
                 \IP^2 & 1 & \piC{\put(1,1){\color{blue}\circle{5}}}2 \\
                 \IP^1 & m & 2{-}m \\}];\\
 &&
 &~~\text{and}&\quad
 X'''_m
 &\in\K[{c||c|c}{\IP^1 & 0 & 2 \\
                 \IP^1 & 0 & 2 \\[1pt]
                  \hdashline[1pt/1pt]\rule{0pt}{12pt}
                 \IP^2 & 1 & \piC{\put(1,1){\color{blue}\circle{5}}}2 \\
                 \IP^1 & m & 2{-}m \\}].
\end{alignat}
The circles highlight the particular degree necessary for the periodicity of the Hirzebruch $n$-fold (appearing below the dashed horizontal line in three of the examples) to be inherited by the Calabi-Yau 3-fold. In turn, the example~\eqref{e:SeqT} does not satisfy this condition, the sequence therein terminates and exhibits none of the periodicity of the Hirzebruch surface in in which those $\Tw{X}_m$ are embedded.

This regularity persists generally, throughout the examples constructed from Hirzebruch $n$-folds, as demonstrated by several more complicated examples in Appendix C.

\subsection{Discrete Deformations and Extremal Transitions}
 \label{s:DDefo}
As  detailed in Appendix~\ref{s:DDa}, it is known that Hirzebruch surfaces of the same homotopy type, $\fF_m\approx\fF_{m+2}$, may be regarded as discrete deformations of one another~\cite{rGHSAR,rBeast}. The direct computations in Appendix~\ref{s:HFm} are consistent with our conjecture~\ref{c:DDeFm}, that the same is true of the straightforward higher-dimensional generalizations,
\begin{equation}
   \sF^{\sss(n)}_m \in \K[{c||c}{\IP^n&1\\\IP^1&m}],\qquad
   2\leqslant n\in\ZZ~~\text{and}~~0\leqslant m\in\ZZ.
\end{equation}
It therefore seems natural to propose:
\begin{conj}\label{c:DDeXm}
({\bfseries i})~The deformation spaces of Calabi-Yau 3-folds $X_m$ and $X_{m+n}$ which belong to an $[m\pMod{n}]$-periodic sequence of configurations the periodicity of which stems from the same periodicity of a Hirzebruch $n$-fold factor in the embedding space are ``separate but infinitesimally near,'' so that $X_m$ is a discrete deformation of $X_{m+n}$.

({\bfseries ii})~In any classical field theory, the use of $X_m$ and $X_{m+n}$ should produce identical models; however, some quantum effects may well distinguish $X_m$ from $X_{m+n}$; see Section~\ref{s:Coda}.
\end{conj}

\begin{figure}[htbp]
 \begin{center}
  \begin{picture}(140,55)
   \put(0,-5){\includegraphics[width=140mm]{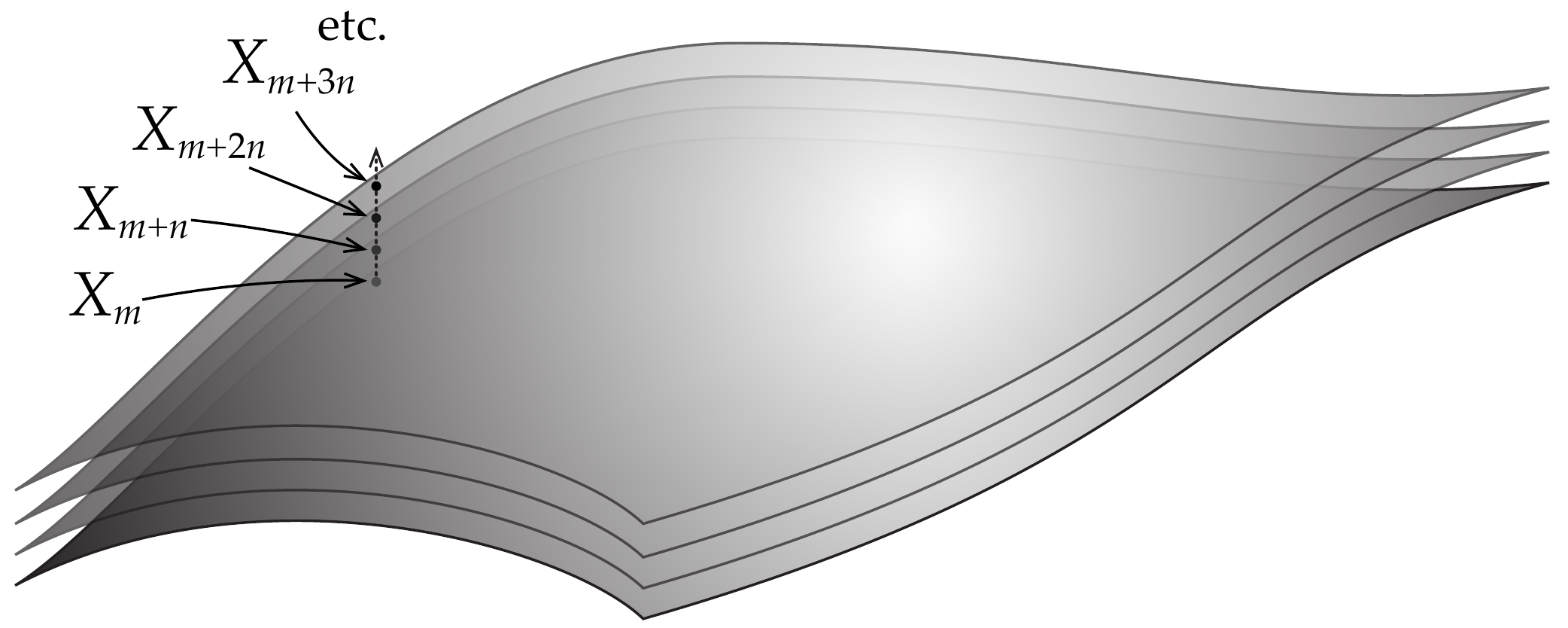}}
  \end{picture}
 \end{center}
 \caption{The Calabi-Yau $(n{-}1)$-folds $X_{m+kn}\subset\sF^{\sss(n)}_{m+kn}$ that are diffeomorphic to each other for $k=0,1,2,\ldots$ have deformation spaces that are infinitesimally close; see Conjecture~\ref{c:DDeXm}.}
 \label{f:DD}
\end{figure}

In particular, the outward emanating sequences of configurations in Figure~\ref{f:PinWheel}, of which the upper right-hand side quarter ($X_0,X_4,X_8,X_{12}\ldots$) is reproduced in~\eqref{e:m=0}, are in fact sequences of such discrete deformations; see Figure~\ref{f:DD}.
The {\em\/local\/} Kodaira-Spencer deformation spaces $H^1(X_m,T)$, $H^1(X_{m+n},T)$, $H^1(X_{m+2n},T)$, $H^1(X_{m+3n},T)$ etc., are of course all isomorphic. Whether this isomorphism extends to the entire moduli spaces as suggested in Figure~\ref{f:DD}, to the cohomology rings defined by the Yukawa couplings, and also away from the ``large radius limit,'' remains an open question. As the Calabi-Yau 3-folds $X_{m+kn}$ are all diffeomorphic for $k=0,1,2,\ldots$ and so represent the same real manifold, the situation in Figure~\ref{f:DD} would imply that the complex structure moduli space of such real manifolds comes in disjoint ``sheets,'' possibly distinguishable by quantum effects as per Conjecture~\ref{c:DDeXm}; see also Section~\ref{s:Coda}.

The first three models in the sequence~\eqref{e:Seq}, $X_0$, $X_1$ and $X_3$, are in fact ``ordinary'' CICYs~\cite{rGHCYCI,rCYCI1}, for all of which it has long since been known that they are connected by way of extremal transitions~\cite{rGHC}. That proof does not extend to the $m\geqslant3$ members of the sequence~\eqref{e:Seq}. However, it can be shown that $X_0,\cdots,X_5$ in~\eqref{e:Seq} have representations as hypersurfaces in toric 4-folds~\cite{rKreSka00b,rtCYdB}, at least as far as the topological data are concerned, and that those toric representations of $X_m$ are connected by way of extremal transitions; these and related matters will be discussed elsewhere.

\section{Summary and Outlook}
\label{s:Coda}
The gCICYs introduced in~\cite{rgCICY1} and further studied in the present paper are providing a promising new class of Calabi-Yau manifolds which extends beyond the current complete intersection (and hypersurfaces) in Fano toric varieties. The novel construction allows for interesting new K3 and elliptically fibered Calabi-Yau manifolds to be found which are important in string duality scenarios and F-theory considerations. In particular, in that context, the perturbative (in $\alpha'$) $d=4$, $N=2$ string vacua, with (often) a dual heterotic $\text{K3}\times T^2$ compactification with some choice of $SU(2)$ instanton embedding in the vector bundle~\cite{rKachru:1995wm}, provide equivalent, perturbative, low-energy effective field theories in terms of several classically isomorphic K3-fibered Calabi-Yau manifolds. However, non-perturbative effects, in particular the world-sheet instantons on the Calabi-Yau 3-folds, may in fact be different and hence provide different non-perturbative completions. 

This issue was studied for hypersurfaces in toric varieties already more than twenty years ago~\cite{rBKK}; see also \cite{rLouis:1996mt}.
Several examples were found where the K\"ahler moduli space has a large radius limit which is a K3-fibration with the Hodge numbers $(2,86)$ and the same topological data as the $m=0,1,2,3,4,5$ gCICY models presented here~\cite{rBKK}. For $m=3,4,5$ the extended K\"ahler moduli space has multiple large radius phases in which the K3-fibered Calabi-Yau phase is obtained by a novel flop~~\cite{rBKK,rBHprime}.
With the mod~4 periodicity of the current work we may then be able to test our conjecture by comparing the Gromow-Witten (GW) invariants that can be calculated for the above mod~4 related Calabi-Yau manifolds. Specifically, let us consider the case of $m\,{=}\,0\pMod4$ realized as hypersurfaces in toric varities. For $m=0$ the Calabi-Yau hypersurface has identical GW invariants to the $m\,{=}\,0$ gCICY, and hence also exhibits a conifold transition to $[\IP^4_{(1,1,1,1,4)}||8]$, see Figure~\ref{f:PinWheel}.\footnote{Note that while we do not have a GLSM formulation for the gCICY construction in general, the existence of hypersurfaces in toric varities with the same topological data as the $X_{3,4,5}$ models indicates that it may be possible to find a generalized GLSM for the gCICYs as well.}
However, the $m=4$ Calabi-Yau hypersurface has GW invariants which differ from those of its mod~4 cousin apart from the invariants associated to the identical K3-fiber~\cite{rBHprime}. In particular, there is a conifold transition to a different $h_{1,1}=1$ Calabi-Yau 3-fold, to $[\IP^4_{(1,1,1,1,2)}||6]$.

This phenomenon extends beyond the particular K3-fibration we have focused on in this paper. Consider the heterotic string compactified on $T^2\times {\rm K3}$ with $SU(2)$ instanton embedding $(4,10,10)$ in the $SU(2)\times E_8\times E_8$ gauge bundle at the $SU(2)$ symmetric point of the $T^2$, which was conjectured to be dual to type IIA theory on the K3-fibered Calabi-Yau hypersurface $[\IP^4_{(1,1,2,2,6)}||12]$ with Hodge numbers $(2,128)$~\cite{rKachru:1995wm}. However,  there are in fact multiple Calabi-Yau three-folds with Hodge numbers $(2,128)$~\cite{rKS-CY,rKreuzer:2000xy}, with instances of an extended moduli space with several Calabi-Yau phases~\cite{rBKK,rBHprime}, analogous to our earlier discussion of the manifolds with Hodge numbers $(2,86)$.
In this case it can be shown that there is a mod~3 periodicity and we once more have multiple diffeomorphic representatives with the same classical topological data, but where the GW invariants differ even after the integral change of basis~\cite{rBHprime}.

Thus, because these latter type IIA vacua do have heterotic duals we then have several different non-perturbative completions of the same perturbative heterotic vacuum. The interesting issue of how this can be understood from the heterotic perspective is left for future investigations.

\acknowledgments 
 We would like to thank Lara Anderson, Charles Doran, James Gray and Richard Wentworth for helpful discussions on  topics in this article, Sheldon Katz and Albrecht Klemm for comments on the paper, and in particular Boris Pioline for collaborating in the early stages of this project.
 P.B. would like to thank the CERN Theory Group for their hospitality over the past several years.
 T.H. is grateful to the Department of Physics, University of Maryland, College Park MD, Department of Physics, University of Central Florida, Orlando FL and the Physics Department of the Faculty of Natural Sciences of the University of Novi Sad, Serbia, for the recurring hospitality and resources.

\appendix
\section{Hirzebruch $n$-folds}
\label{s:HFm}
We compute various useful properties of the 4-folds $F_m$ appearing in~\eqref{e:Seq}, and then discuss their analogues in different dimensions.

\subsection{Anticanonical sections}
\label{s:-K}
As the configuration~\eqref{e:Seq} embeds the Calabi-Yau 3-folds $X_m$ as hypersurfaces in the 4-folds $F_m$, it is imperative to prove that the anticanonical bundle of $F_m$ does have holomorphic sections from which to construct the defining equation of $X_m$.

\subsubsection{Counting and tensor structure}
The anticanonical bundle of the hypersurface $F_m\in\ssK{[}{c||c}{\IP^4 & 1 \\ \IP^1 & m \\}{]}$ is $\cK^*_{F_m}=\cQ=\cO\pM{4\\\!2-m\!\\}$. To determine $H^*(F_m,\cQ)$, we tensor the monad~\eqref{e:KFm} by $\cO\pM{4\\\!2-m\!\\}$ and obtain the Koszul resolution given in the header row of the tabulation of the so-valued cohomology:
\begin{equation}
   \begin{array}{c|ccccl}
  &\cO_{\!A}\pM{3\\\!2-2m\!\\} &\overset{p}\into &\cQ=\cO_{\!A}\pM{4\\\!2-m\!\\}
   &\overset{\rho_F}\onto &\cQ|_{F_m} \\*[1mm] \hline \rule{0pt}{6mm}
 0. &\q_m^1\{\vf_{(abc)(i_1\cdots\,i_{2-2m})}\}
     &\too{p} &\q_m^2\{\f_{(abcd)(i_1\cdots\,i_{2-m})}\}
      &\too{\rho_F}&H^0(F_m,\cQ)\too{\rd} \\*[2mm]
 1. &\q_2^m\{\ve_{\9}^{i(j}\vf_{(abc)}^{k_1\cdots\,k_{2m-4})}\}
     &\too{p} &\q_4^m\{\ve_{\9}^{i(j}\f_{(abcd)}^{k_1\cdots\,k_{m-4})}\}
      &\too{\rho_F}&H^1(F_m,\cQ)\too{\rd} \\*
 2. &0&&0&&H^2(F_m,\cQ)=0 \\*[-2mm]
 \vdots &\vdots&&\vdots&&~~~\vdots \\*
  \end{array}
 \label{e:mH*Q}
\end{equation}
The ``direct image'' tensor representatives~\cite{rBE,rBeast} of the so-valued cohomology groups are tabulated underneath the corresponding sheaves.
 The appearances of the step-function
\begin{equation}
  \q_m^n = \bigg\{\begin{array}{rl}1&m\leqslant n,\\[2pt]0&m>n.\\\end{array}
\end{equation}
indicate that there are four separate cases:

\paragraph{\boldmath$m=0,1$:} All the contributions are in the top, $0^\text{th}$ cohomology row, and produce 105 equivalence classes of polynomials (see~\eqref{e:4dFmK*} below):
\begin{alignat}9
  H^0(F_m,\cQ)
  &= \big\{(\f_{(abcd)(ij)}\,/\,p_{(a}f_{bcd)(ij)})\,x^ax^bx^cx^d\,y^iy^j\big\},
   &\quad m&=0;\\
  &= \big\{(\f_{(abcd)\,i}\,/\,f_{(abc}\,p_{d)\,i})\,x^ax^bx^cx^d\,y^i\big\},
   &\quad m&=1.
 \label{e:-4K01}
\end{alignat}

\paragraph{\boldmath$m=2$:} There are now two separate contributions,
 $H^0\big(A,\cO\pM{4\\0\\}\big)=\{\f_{(abcd)}\}$ in the middle of the $0^\text{th}$ cohomology row and
 $H^1\big(A,\cO\pM{~\>3\\-2\\}\big)=\{\ve_{\9}^{ij}f_{(abc)}\}$ on the left of the $1^\text{st}$ cohomology row. This results in:
\begin{subequations}
\begin{gather}
 0\to H^0(A,\cQ)\too{\rho_F} H^0(F_2,\cQ)\too{\rd} H^1\big(A,\cO\pM{~\>3\\-2\\}\big)\to0,\\
  H^0(F_2,\cQ)
   = \big\{\big(\f_{(abcd)} +
                \g_{(abcd)\,k}^{\,i}\,{\ttt\frac{y^k}{y^j}}\big)\,
        x^ax^bx^cx^d\big\},\quad
        \g_{(abcd)\,k}^{\,i}\define\ve_{\9}^{ij}f_{(abc}\,p_{d)(jk)}\label{e:2H0}
\end{gather}
\end{subequations}
The Laurent polynomial $\g_{(abcd)\,k}^{\,i}\,x^ax^bx^cx^d\,\frac{y^k}{y^j}$ is one of the equivalent representatives generated by the defining equation~\eqref{e:fDef} of $F_2$. As shown explicitly by~\eqref{e:2gy1} and~\eqref{e:2gy0} below, this $F_2$-equivalence class of 35 holomorphic sections of $\cQ|_{F_2}$ contains representatives that are well-defined over every point of $\IP^1$.

\paragraph{\boldmath$m=3$:} The only nonzero contribution is now $H^1\big(A,\cO\pM{~\,3\\\!-4}\big)$, in the second row of the left column in~\eqref{e:mH*Q}, producing:
\begin{subequations}
\begin{gather}
  0\to H^0(F_3,\cQ) \too{\rd} H^1\big(A,\cO\pM{~\,3\\\!-4\\}\big)\to0,\\
  H^0(F_3,\cQ)
   =\big\{\g_{(abcd)}^{\,i}{\ttt\frac{x^ax^bx^cx^d}{y^i}}\big\},\quad
  \g_{(abcd)}^{\,i}\define\ve_{\9}^{ij\9}f_{(abc}^{(kl)}\,p^{\9}_{d)(jkl)}.
   \label{e:3H0}
\end{gather}
\end{subequations}
As in~\eqref{e:2g} below, the inclusion of the $\ve^{ij}\,p_{a(jkl)}$ factor and the vanishing of $p(x,y)$ turns these Laurent polynomials into an $F_3$-equivalence class of 105 holomorphic sections of $\cQ|_{F_3}$, with well-defined representatives over every point of $\IP^4\,{\times}\,\IP^1$.

\paragraph{\boldmath$m\geqslant4$:} Now both contributions in the second row in~\eqref{e:mH*Q} are nonzero, and fit into the sequence:
\begin{equation}
 0\to H^0(F_m,\cQ)\too{\rd} H^1\big(A,\cO\pM{3\\\!2-2m\!\\}\big)\too{p}
      H^1\big(A,\cO\pM{4\\\!2-m\!\\}\big) \too{\rho} H^1(F_m,\cQ) \to0.
 \label{e:4H0seq}
\end{equation}
This specifies $H^0(F_m,\cQ)$ as the preimage by the differential $\rd$-map of the ``direct image'' within $H^1\big(A,\cO\pM{3\\\!2-2m\!\\}\big)$:
\begin{subequations}
 \label{e:qxy}
\begin{alignat}9
 \g^{(j_1{\cdots}j_{m-2})}_{(abcd)}
 &\define \ve_{\9}^{i(j_1}\vf^{j_2{\cdots}j_{2m-3})}_{(abc}\,
             p^{\9}_{d)(i\,j_{m-1}{\cdots}j_{2m-3})}, \label{e:4H0}\\
 \text{where}&\quad \vf^{(j_1{\cdots}j_{2m-4})}_{(abc}\,
             p^{\9}_{d)(j_{m-3}{\cdots}j_{2m-4})}=0. \label{e:4qker}
\intertext{which are then used to construct the Laurent polynomials for $H^0(F_m,\cQ)$:}
 \g(x,y)
 &:=\ve_{\9}^{i(j_1}\vf^{j_2{\cdots}j_{2m-3})}_{(abc}\,
      p^{\9}_{d)(i\,j_{m-1}{\cdots}j_{2m-3})}\,
       \frac{x^ax^bx^cx^d}{g^{(j_1{\cdots}j_{m-2})}(y)}. \label{e:4qxy}
\end{alignat}
\end{subequations}
The form of the condition~\eqref{e:4qker} is dictated by the only covariant way to contract the tensor representatives $\ve_{\9}^{ij}f^{(k_1{\cdots}\,k_{2m-4})}_{(abc)}$ of $H^1\big(A,\cO\pM{3\\\!2-2m\!\\}\big)$ with $p_{a(i_1\cdots\,i_m)}$ so as to produce the tensor coefficients of a degree-$\pM{4\\2-m}$ polynomial.
 The $(m{-}1)$ degree-$(m{-}2)$ generic $\IP^1$-polynomials $g^{(j_1{\cdots}j_{m-2})}(y)$ allow separating the poles of $\g(x,y)$ to $(m{-}1)(m{-}2)$ distinct locations and minimally extends the ``direct image'' linear algebra methods~\cite{rBE,rBeast} to accommodate the manifestly non-linear nature of the generalized complete intersections~\eqref{e:Seq} for $m\geqslant3$. It also facilitates using the defining equation~\eqref{e:p=0} of $F_m$ to construct well-defined holomorphic sections~\eqref{e:4qxy} of $\cO\pM{4\\\!2-m\!\\}$ over $F_m$ for every $m\geqslant0$, the~\eqref{e:2g}-like {\em\/equivalence classes\/} of which fully cover the explicit case-by-case constructions of the type given in Ref.~\cite{rgCICY1}.

The $\binom{4+4}4{\cdot}\binom{(m-4)+1}1=70(m{-}3)$ constraints in the system~\eqref{e:4qker} must leave {\em\/at least\/} 105 of the $\binom{3+4}4{\cdot}\binom{(2m-4)+1}1=35(2m{-}3)$ tensor coefficients $\ve_{\9}^{ij}\vf^{(k_1{\cdots}\,k_{2m-4})}_{(abc)}$ free to span $H^0(F_m,\cQ)$. In fact, this is an undercount for $m\geqslant4$, and the exact result is
\begin{equation}
  H^0(F_m,\cK^*)=105+\d^{\sss(4)}_m,\quad
  H^1(F_m,\cK^*)=\d^{\sss(4)}_m \quad\text{and}\quad
  \d^{\sss(4)}_m = \q^m_3\,15(m{-}3).
 \label{e:4dm}
\end{equation}
The computation of $\d^{\sss(4)}_m$ is given in~\eqref{e:nPEP1}--\eqref{e:ndm} below, for general Hirzebruch $n$-folds. Stated differently and for $m\geqslant4$, 105 is the index of the cohomology map generated by multiplication with the defining polynomial $p(x,y)$ in degree-1 row of~\eqref{e:mH*Q}.

To summarize, we have obtained:
\begin{equation}
 \begin{array}{@{}r|l|l|l@{}}
  \boldsymbol{m} &\boldsymbol{H^0(F_m,\cQ),~~\dim F_m=4}
   &\textbf{Number} &\textbf{Sections} \\
 \hline\rule{0pt}{13pt}
  0 & \{\f_{(abcd)(ij)}/p_{(a}\vf_{bcd)(ij)}\}
     &\binom{4+4}4\binom{2+1}1{-}\binom{3+4}4\binom{2+1}1=105
      & \text{ordinary} \\[1mm] \hline \rule{0pt}{12pt}
  1 & \{\f_{(abcd)\,i}/\vf_{(abc}\,p_{d)\,i}\}
     &\binom{4+4}4\binom{1+1}1{-}\binom{3+4}4\binom{0+1}1=105
      & \text{ordinary} \\[1mm] \hline \rule{0pt}{12pt}
 \MR2*2
    & \{\f_{(abcd)}\}
     &\binom{4+4}4\binom{0+1}1=70
      & \text{ordinary} \\
    & \{\ve_{\9}^{ij}\vf_{(abc}\,p_{d)(ik)}\}
     &\binom{3+4}4\binom{0+1}1=35
      & \text{Laurent} \\[1mm] \hline \rule{0pt}{13pt}
  3 & \{\ve_{\9}^{i(j\9}\vf_{(abc}^{kl)}\,p^{\9}_{d)(ikl)}\}
     &\binom{3+4}4\binom{2+1}1=105
      & \text{Laurent} \\[1mm] \hline \rule{0pt}{13pt}
 \MR2*{$\geqslant4$}
    & \{\ve_{\9}^{i(j_1\9}\vf_{(abc}^{j_2{\cdots}j_{2m-3})}\,
                           p^{\9}_{d)(ij_{m-3}{\cdots}j_{2m-3})}\}
     &\binom{3+4}4(2m{-}3)
      & \text{Laurent} \\
    & \vf_{(abc}^{(i_1\cdots\,i_{2m-4})}\,p^{\9}_{d)(i_1\cdots\,i_{2m-4})}=0
     &~~-\binom{4+4}4(m{-}3)\leqslant105+\d^{\sss(4)}_m{}^\ddag &\\[1mm] \hline
\MC4l{\text{\footnotesize$^\ddag$~The ``excess'' number of sections $\d^{\sss(4)}_m = \q^m_3\,15(m{-}3)$ is computed in~\eqref{e:nPEP1}--\eqref{e:ndm}.}}
  \end{array}
 \label{e:4dFmK*}
\end{equation}

\subsubsection{Being well-defined}
That holomorphic functions on $F_m$ are equivalence classes of functions on $A$ modulo $p(x,y)$-multiples of $\cO_{\!A}\pM{\!-1\\\!-m\\}$-valued functions by~\eqref{e:KFm} is crucial in showing that the above-obtained Laurent polynomials are well-defined on $F_m$. Suffice it here to show this for $m=2$: Without loss of generality, we may write the defining equation of $F_2$ as
\begin{equation}
  p(x,y) = p_{00}(x)\,(y^0)^2 +2 p_{01}(x)\,y^0y^1 + p_{11}(x)\,(y^1)^2 ~=0.
 \label{e:p=0}
\end{equation}
In turn, the second, $\g^i_{(abcd)\,k}$-parametrized term in~\eqref{e:2H0} results in the Laurent polynomial
\begin{subequations}
 \label{e:2g}
\begin{alignat}9
  \g(x,y)
  &=\ve_{\9}^{ij}\vf_{(abc}\,p_{d)(ik)}\, x^ax^bx^cx^d\,\frac{y^k}{y^j}
   =\vf(x)\Big(p_{00}(x)\frac{y^0}{y^1}-p_{11}(x)\frac{y^1}{y^0}\Big).
 \label{e:2ef}
\intertext{The vanishing~\eqref{e:p=0} of $p(x,y)$ on $F_2$ implies that this is equivalent to:}
  \g(x,y)
  &=\vf(x)\bigg[p_{00}(x)\frac{y^0}{y^1} -p_{11}(x)\frac{y^1}{y^0}
    +\l\Big(\underbrace{p_{00}(x)\,\frac{y^0}{y^1} +2 p_{01}(x)
                       +p_{11}(x)\,\frac{y^1}{y^0}}_{=0~\text{on $F_2$ owing to~\eqref{e:p=0}}}\Big)\bigg].
 \label{e:gxy}
\intertext{In particular, this $\l$-{\em\/continuum\/} of equivalent representatives includes:}
 &\overset{(\ref{e:p=0})}\simeq
  -2\,\vf(x)\Big(p_{01}(x) +p_{11}(x)\,\frac{y^1}{y^0}\Big),\quad
 \text{at}~y^0\neq0,~\text{choose}~\l\to-1; \label{e:2gy1}\\
 &\overset{(\ref{e:p=0})}\simeq
  +2\,\vf(x)\Big(p_{01}(x) +p_{00}(x)\,\frac{y^0}{y^1}\Big),\quad
 \text{at}~y^1\neq0,~\text{choose}~\l\to+1, \label{e:2gy0}
\end{alignat}
\end{subequations}
and which are holomorphic in the indicated regions. This {\em\/equivalence class\/} of degree-$\pM{4\\0\\}$ rational polynomials over $\IP^4\,{\times}\,\IP^1$ then provides sections of $\cO\pM{4\\\!2-m\!\\}$ that are well-defined and holomorphic everywhere on $F_2$.
 This irresistibly reminds of the well-known Wu-Yang construction of the magnetic monopole, since:
 ({\small\bf1})~neither of the expressions~\eqref{e:2g} is well-defined everywhere on $\IP^1$,
 ({\small\bf2})~at least one of~\eqref{e:2g} is well-defined everywhere on $\IP^1$,
 ({\small\bf3})~wherever on $\IP^1$ that both of~\eqref{e:2gy1} and~\eqref{e:2gy0} are well-defined, they are equivalent owing to~\eqref{e:p=0}.
 Together with the ``ordinary'' $\IP^4$-quadrics $\f(x)=\f_{abcd}\,x^ax^bx^cx^d$, the Laurent polynomials $\g(x,y)$ provide $\binom{4+3}3+\binom{3+3}3=70+35=105$ sections for $H^0(F_m,\cQ)$ with which to define Calabi-Yau 3-folds $X_2\subset F_2$.

Conversely, the degree-$\pM{3\\\!2-2m\!\\}$ Laurent polynomials $\vf(x,y)$ which in~\eqref{e:4qxy} parametrize the anticanonical sections $\g(x,y)|_{F_m}\in H^0(F_m,\cQ)$ are localized to the hypersurface 4-fold $F_m\define \{(x,y)\in\IP^4\,{\times}\,\IP^1: p(x,y)=0\}$ by means of the residue formula~\cite{rRes}:
\begin{subequations}
 \label{e:qRes}
\begin{alignat}9
  \vf(x,y)|_{F_m}
   &\define\oint_{\Gamma(F_m)} \frac{(y\,\rd y)}{p(x,y)}~\g(x,y)
    =\W^{(i_1\cdots\,i_{m-2})}(x)\big[\vd_{i_1}\!\cdots\vd_{i_{m-2}}\g(x,y)\big]_{F_m},\\
  \W^{(i_1\cdots\,i_{m-2})}(x)&\define\oint_{\Gamma(F_m)}
       \frac{(y\,\rd y)}{[\vd_{i_1}\cdots\vd_{i_{m-2}}p(x,y)]}.
\end{alignat}
\end{subequations}
Here, $\Gamma(F_m)$ is the $(S^1\,{\times}\,F_m)$-like ``Gaussian'' boundary of a sufficiently ``thin'' tubular neighborhood of $F_m\subset\IP^4\,{\times}\,\IP^1$, $\W^{(i_1\cdots\,i_{m-2})}(x)$ are degree-$\pM{\!-1\\~\,0\\}$ holomorphic $\IP^1$-constant 0-forms on $F_m$. This type of residue formula has been shown to represent the cohomology elements in all complete intersections in (even weighted) projective spaces~\cite{rRes}, and it is gratifying to find that~\eqref{e:qRes} also extends to the generalized complete intersections of Ref.~\cite{rgCICY1}.

\subsection{Other properties of $F_m$}
\label{s:OtherFm}
The computation~\eqref{e:mH*Q} may be generalized to produce the {\em\/plurigenera\/}
\begin{equation}
   \fP_{-k}(F_m)
    \define \sum_{r=0}^4(-1)^r \dim H^r\big(F_m,(\cK_{F_m}^*)^{\otimes k}\big)
   ={\ttt\frac13}(2k+1)^2 (4k+1) (4k+3),
 \label{e:Pg}
\end{equation}
where $\cK_{F_m}^*=\cO\pM{4\\\!2-m\!\\}\cQ$, and which is independent of $m$ for all $k$; also, $\fP_{-k}=\fP_{k+1}$.

The Lefschetz hyperplane theorem~\cite{rB+S-Adjunction,rBeast} is applicable to $F_m$ for all $m>0$, while $F_0=\IP^3\,{\times}\,\IP^1$ straightforwardly.
 Together with Poincar\'e duality, the K\"unneth formula  and the universal coefficient theorem~\cite{rMcC-SS}, this guarantees~\cite[p.\,44]{rBeast} that $H^r(F_m,\ZZ)\approx H^r(\IP^4\,{\times}\,\IP^1,\ZZ)$ for all $r\neq4$. Moreover, the surjection $H^4(F_m,\ZZ)\onto H^4(\IP^4\,{\times}\,\IP^1,\ZZ)$ is in fact an isomorphism since $\chi_{_E}(F_m)=8$, so that $H^*(F_m,\ZZ)=H^*(\IP^4\,{\times}\,\IP^1,\ZZ)$ and with no torsion. In particular, $H^{1,1}(F_m)\approx H^2(F_m,\ZZ)\approx H^2(\IP^4\,{\times}\,\IP^1,\ZZ)$ is generated by the pull-backs of the K\"ahler forms of $\IP^4$ and $\IP^1$---which are thus guaranteed to generate the Chern class of $F_m$ for all $m\geqslant0$. This agrees with the direct computation using the adjunction relation for $F_m$:
\begin{equation}
 \begin{array}{rccccc}
\MR4*{$F_m$-resolution $\left\{\rule{0pt}{30pt}\right.$}
 &  (\cQ^*_A)^{\oplus2} &\too{\rd p} &T_A\otimes\cQ^*_A\\
 &\WC{\into}\raisebox{2pt}{$\SSS\,p$} &&\WC{\into}\raisebox{2pt}{$\SSS\,p$} \\[-1mm]
 &  \cQ^*_A &\too{\rd p} &T_A\\
 &\WC{\onto}\raisebox{2pt}{$\SSS\rho$} &&\WC{\onto}\raisebox{2pt}{$\SSS\rho$} \\[-2mm]
\text{dual adjunction}:~~
 &  \cQ^*|_{F_m} &\overset{\rd p}{\into} &T^*_A|_{F_m} &\onto &T^*_{F_m}
 \end{array}
 \label{e:KrAdj}
\end{equation}

Having determined that (the pullbacks of) the K\"ahler forms $J_1$ of $\IP^n$ and $J_2$ of $\IP^1$ generate $H^{1,1}(F_m)\cap H^2(F_m,\ZZ)$ for all $m\geqslant0$, the straightforward Chern class, the intersection and various Chern evaluation computations produce:
\begin{subequations}
 \label{e:Fmod4c}
\begin{gather}
  \begin{array}{rl}
  c&=\big(4J_1 + (2{-}m)J_2\big)
     +\big(6J_1^{~2} + (8{-}3m)J_1 J_2\big)\\
   &~\quad+\big(4J_1^{~3} + (12{-}3m)J_1^{~2}J_2\big)
     +\big(J_1^{~4} + (8{-}m)J_1^{~3} J_2\big),
  \end{array}\\
  C_1^{~4} =512,\qquad
  C_1^{~2}{\cdot}C_2 =224,\qquad
  C_1{\cdot}C_3 =56,\qquad
  C_2^{~2} =96,\qquad
  C_4 = \chi_{_E} =8, \label{e:FmCc}\\
  C_1^{~3}[aJ_1{+}bJ_2] =16[6a + (\2{4b{+}am})],\qquad
  C_1{\cdot}C_2[aJ_1{+}bJ_2] =2[22a + 3(\2{4b{+}am})],\\
  C_3[aJ_1{+}bJ_2] =12a + (\2{4b{+}am}),\\
  C_1^{~2}[(aJ_1{+}bJ_2)^2] =8a[2a+(\2{4b{+}am})],\qquad
  C_2[(aJ_1{+}bJ_2)^2] =a(8a+3(\2{4b{+}am})],\\
  C_1[(aJ_1{+}bJ_2)^3] =a^2(2a +3(\2{4b{+}am})],\qquad
  [(aJ_1{+}bJ_2)^4]_{F_m} =a^3(\2{4b{+}am}).
\end{gather}
\end{subequations}
All the Chern numbers are $m$-independent~\eqref{e:FmCc}, and all the various Chern  evaluations on $H^{1,1}(F_m)$ depend on $m$ and $bJ_2$ only through the (underlined) combination $(4b{+}ma)$. This indicates an $[m\pMod4]$-relation:
\begin{equation}
   F_m \approx F_{m+c}:\qquad
   \bM{J_1\\J_2\\}_m \overset{\approx}{\longleftrightarrow}
  \bM{1&~-\frac{c}4\\[2pt]0&~~~1\\[2pt]}\bM{J_1\\J_2\\}_{m+c}\quad \text{iff}~~c\in4\ZZ.
 \label{e:Fmod4}
\end{equation}
The {\em\/differences\/} in the topological data~\eqref{e:Fmod4c} insure that there are {\em\/at least\/} four distinct diffeomorphism classes.
 While we are not aware of a 4-fold classification result as straightforwardly precise as Wall's theorem~\cite{rWall,rBeast} that classifies the diffeomorphism class of Calabi-Yau 3-folds, we will assume that the relation~\eqref{e:Fmod4} between $F_m \approx F_{m+4}$ is in fact a diffeomorphism.
 That is, we assume that the above topological data insures that the sequence of Hirzebruch 4-folds $F_m$ forms precisely four diffeomorphism classes, $[F_m]\approx F_{m\pMod4}$ for $0\leqslant m\in\ZZ$.
 The first two of these four diffeomorphism classes each have a Fano representative: $F_0=\IP^3\,{\times}\,\IP^1$ and $F_1=\ssK[{c||c}{\IP^4&1\\\IP^1&1\\}]$, while the anticanonical bundle of $F_2$ and $F_3$ are evidently non-positive over $\IP^1$ and are not Fano. The two-dimensional analogues of these results are well-known~\cite{rGrHa}; see below.

\subsection{Other dimensions}
\label{s:dFm}
Analogously to the sequence~\eqref{e:Seq}, sequences~\eqref{e:3dF:Xm} and~\eqref{e:2dF:Xm} involve 3- and 2-dimensional analogues of the 4-fold $F_m$. Listing them side-by-side,
\begin{equation}
  \fF_m\in\ssK{[}{c||c}{\IP^2 & 1 \\ \IP^1 & m \\}{]},\qquad
  \cF_m\in\ssK{[}{c||c}{\IP^3 & 1 \\ \IP^1 & m \\}{]},\qquad
    F_m\in\ssK{[}{c||c}{\IP^4 & 1 \\ \IP^1 & m \\}{]},\quad\cdots\quad
  \sF^{\sss(n)}_m\in\ssK{[}{c||c}{\IP^n & 1 \\ \IP^1 & m \\}{]},
 \label{e:dFm}
\end{equation}
make it obvious that these degree-$\pM{1\\m\\}$ hypersurfaces in $\IP^n\,{\times}\,\IP^1$ are at every point of $\IP^1$ simple hyperplanes in $\IP^n$, i.e., $[\IP^n||1]\approx\IP^{n-1}$. Varying then the base-point over $\IP^1$, each such hypersurface forms an $m$-twisted $\IP^{n-1}$-bundle over $\IP^1$.
It is worthwhile noting that the $n=1$-dimensional case of such varieties,
\begin{equation}
  \mathfrak{f}_m \in\ssK{[}{c||c}{\IP^1_x & 1 \\[1pt] \IP^1_y & m \\}{]}
 \label{e:1dFm}
\end{equation}
are $m$-twisted 1-point fibrations over (simple covers of) $\IP^1_y$, where the 1-point fiber is the hyperplane $[\IP^1_x||1]$. Alternatively, they may also be understood ``the other way around,'' as an $m$-fold ramified cover of $\IP^1_x$: at each point $x_*\in\IP^1_x$, the defining equation $p(x_*,y)$ is a degree-$m$ polynomial over $\IP^1_y$. The zero-locus of this degree-$m$ polynomial consists of $m$ $\IP^1_y$-points, thus producing an $m$-fold cover of $\IP^1_x$, ramified (branched) at the $\IP^1_y$-locations where the zeros of $p(x,y)$ coalesce.

\paragraph{Hodge numbers:}
Just as for the 4-fold $F_m$ above, the Lefschetz hyperplane theorem applies for all $m,n>0$, while $\sF^{\sss(n)}_0=\IP^{n-1}\,{\times}\,\IP^1$, straightforwardly. Together with Poincar\'e duality, the K\"unneth formula  and the universal coefficient theorem~\cite{rMcC-SS}, this guarantees that $H^r(\sF^{\sss(n)}_m,\ZZ)\approx H^r(\IP^n\,{\times}\,\IP^1,\ZZ)$ for all $r\neq n$, and $H^n(\sF^{\sss(n)}_m,\ZZ)\onto H^n(\IP^n\,{\times}\,\IP^1,\ZZ)$ is an isomorphism precisely if $h^{n,n}(\IP^4\,{\times}\,\IP^1)=h^{n,n}(\sF)$. This last condition is in turn guaranteed by the standard computation of $\chi_{_E}(\sF^{\sss(n)}_m)=2n$, whereby $H^*(\sF^{\sss(n)}_m,\ZZ)=H^*(\IP^4\,{\times}\,\IP^1,\ZZ)$ and with no torsion.
 In particular, $H^{1,1}(\sF^{\sss(n)}_m)\approx H^2(\sF^{\sss(n)}_m,\ZZ)\approx H^2(\IP^n\,{\times}\,\IP^1,\ZZ)$, and is generated by the pull-backs of the K\"ahler forms of $\IP^n$ and $\IP^1$---which are thus guaranteed to generate the Chern class of $\sF^{\sss(n)}_m$---for all $n\geqslant2$ and all $m\geqslant0$. As above, this result may  be verified by direct computation for all $n\geqslant2$ using (the dual of) the ($F_m$-resolved adjunction) relation~\eqref{e:KrAdj}.

\paragraph{Chern and intersection numbers:}
Just as was done for the 4-fold $F_m$ above, we readily compute the Chern numbers and the various Chern evaluations for all $n\geqslant2$. For the 2-fold $\fF_m$ we compute:
\begin{subequations}
 \label{e:Fmod2c}
\begin{gather}
  c=\big(2J_1 + (2{-}m)J_2\big) +\big(J_1^{~2} + (4{-}m)J_1 J_2\big),\qquad
  C_1^{~2} =8,\qquad
  C_2 = \chi_{_E} =4,\\
  C_1[aJ_1{+}bJ_2] =2a+(\2{2b{+}am}),\qquad
  [(aJ_1{+}bJ_2)^2]_{\fF_m} =a(\2{2b{+}am}).
\end{gather}
\end{subequations}
The tandem of facts:
 ({\small\bf1})~the Chern numbers are $m$-independent, and
 ({\small\bf2})~the intersections and Chern evaluations depend on $bJ_2$ and $m$ only through the (underlined) hallmark combination $(2b{+}am)$,
 demonstrates the known homotopy type $[m\pMod2]$-periodicity of Hirzebruch surfaces:
\begin{equation}
   \fF_m \approx \fF_{m+c}:\qquad
   \bM{J_1\\J_2\\}_m \overset{\approx}{\longleftrightarrow}
  \bM{1&~-\frac{c}2\\[2pt]0&~~~1\\[2pt]}\bM{J_1\\J_2\\}_{m+c}\quad \text{iff}~~c\in2\ZZ.
 \label{e:Fmod2}
\end{equation}
That is, the sequence of Hirzebruch surfaces $\fF_m$ forms two diffeomorphism classes, $[\fF_{2k}]\approx\fF_0$ and $[\fF_{2k+1}]\approx\fF_1$, for $0\leqslant k\in\ZZ$; both of these have a Fano (del Pezzo) representative: $\fF_0=\IP^1\,{\times}\,\IP^1$ and $\fF_1=\ssK[{c||c}{\IP^2&1\\\IP^1&1\\}]$.\footnote{All Hirzebruch surface $\fF_1$ is {\em\/rigid\/}, i.e., has no deformations~\cite{rBeast}. The deformation family represented by the configuration $\ssK[{c||c}{\IP^2&1\\\IP^1&1\\}]$ thus consists of a single member, so that writing ``='' instead of ``$\in$'' is justified.}

\bigskip
Similarly, for the 3-fold $\cF_m$ we compute:
\begin{subequations}
 \label{e:Fmod3c}
\begin{gather}
  c =\big(3J_1 + (2{-}m) J_2\big)
     +\big(3J_1^{~2} + (6{-}2m) J_1 J_2\big)
     +\big(J_1^{~3} + (6{-}m) J_1^{~2} J_2,\\
  C_1^{~3} =54,\qquad
  C_1{\cdot}C_2 =24,\qquad
  C_3 = \chi_{_E} =6,\\
  C_1^{~2}[aJ_1{+}bJ_2] =12a + 3(\2{3b{+}am}),\qquad
  C_2[aJ_1{+}bJ_2] =6a + (\2{3b{+}am}),\\
  C_1[(aJ_1{+}bJ_2)^2] =2a^2 + 2a(\2{3b{+}am}),\qquad
  [(aJ_1{+}bJ_2)^3]_{\cF_m} =a^2(\2{3b{+}am}).
\end{gather}
\end{subequations}
Again, the tandem of facts:
 ({\small\bf1})~the Chern numbers are $m$-independent, and
 ({\small\bf2})~the intersections and Chern evaluations depend on $bJ_2$ and $m$ only through the (underlined) hallmark combination $(3b{+}am)$,
 demonstrates the homotopy type $[m\pMod3]$-periodicity of Hirzebruch 3-folds\footnote{For 3-folds, Wall's theorem~\cite{rWall,rBeast} does imply that the relationship in~\eqref{e:Fmod3} is a diffeomorphism.}:
\begin{equation}
   \cF_m \approx \cF_{m+3c}:\qquad
   \bM{J_1\\J_2\\}_m \overset{\approx}{\longleftrightarrow}
  \bM{1&-c\\[2pt]0&~~1\\[2pt]}\bM{J_1\\J_2\\}_{m+2c},\quad c\in\ZZ.
 \label{e:Fmod3}
\end{equation}
That is, the sequence of Hirzebruch 3-folds $\cF_m$ forms three diffeomorphism classes, $[\cF_{3k}]\approx\cF_0$, $[\cF_{3k+1}]\approx\cF_1$ and $[\cF_{3k+2}]\approx\cF_2$, for $0\leqslant k\in\ZZ$; the first two of these have a Fano representative: $\cF_0=\IP^2\,{\times}\,\IP^1$ and $\cF_1=\ssK[{c||c}{\IP^3&1\\\IP^1&1\\}]$, while the anticanonical bundle of $\cF_2$ is non-positive over $\IP^1$, as computed explicitly in~\eqref{e:3dFmK*}, below.

\paragraph{Anticanonical sections:}
The computation~\eqref{e:mH*Q} easily adapts to all $n$: the anticanonical bundle becomes a restriction of $\cK^*_{\fF_m}=\cO\pM{n\\\!2-m\!\\}$, and we have:
\begin{equation}
   \begin{array}{c|ccccl}
 n\geqslant2 &\cO_{\!A}\pM{n-1\\\!2-2m\!\\} &\overset{p}\into &\cO_{\!A}\pM{n\\\!2-m\!\\}
   &\overset{\rho_\fF}\onto &\cK^*_{\sF^{\sss(n)}_m} \\*[1mm] \hline \rule{0pt}{6mm}
 0. &\q_m^1\{f_{(a_1\cdots\,a_{n-1}\,(i_1\cdots\,i_{2-2m})}\}
     &\too{p} &\q_m^2\{\f_{(a_1\cdots\,a_n)(i_1\cdots\,i_{2-m})}\}
      &\too{\rho_\sF}&H^0(\sF^{\sss(n)}_m,\cK^*)\too{\rd} \\*[2mm]
 1. &\q_2^m\{\ve_{\9}^{i(j}f_{(a_1\cdots\,a_{n-1})}^{k_1\cdots\,k_{2m-4})}\}
     &\too{p} &\q_4^m\{\ve_{\9}^{i(j}\f_{(a_1\cdots\,a_n)}^{k_1\cdots\,k_{m-4})}\}
      &\too{\rho_\sF}&H^1(\sF^{\sss(n)}_m,\cK^*)\too{\rd} \\*
 2. &0&&0&&H^2(\sF^{\sss(n)}_m,\cK^*)=0 \\*
 \vdots~&\vdots&&\vdots&&~~\vdots \\*[-1mm]
  \end{array}
 \label{e:nmH*Q}
\end{equation}
Akin to the 4-fold $F_m$ case~\eqref{e:4dFmK*}, this produces for the familiar ($n=2$) Hirzebruch surfaces:
\begin{equation}
 \begin{array}{@{}r|l|l|l@{}}
  \boldsymbol{m} &\boldsymbol{H^0(\fF_m,\cK^*),~~\dim\fF_m=2}
   &\textbf{Number} &\textbf{Sections} \\
 \hline\rule{0pt}{13pt}
  0 & \{\f_{(ab)(ij)}/p_{(a}\vf_{b)(ij)}\}
     &\binom{2+2}2\binom{2+1}1-\binom{1+2}2\binom{2+1}1=9
      & \text{ordinary} \\[1mm] \hline \rule{0pt}{12pt}
  1 & \{\f_{(ab)\,i}/\vf_{(a}\,p_{b)\,i}\}
     &\binom{2+2}2\binom{1+1}1-\binom{1+2}2\binom{0+1}1=9
      & \text{ordinary} \\[1mm] \hline \rule{0pt}{12pt}
 \MR2*2
    & \{\f_{(ab)\,i}\}
     &\binom{2+2}2\binom{0+1}1=6
      & \text{ordinary} \\
    & \{\ve_{\9}^{ij}\vf_{(a}\,p_{b)(ik)}\}
     &\binom{1+2}2\binom{0+1}1=3
      & \text{Laurent} \\[1mm] \hline \rule{0pt}{13pt}
  3 & \{\ve_{\9}^{i(j\9}\vf_{(a}^{kl)}\,p^{\9}_{b)(ikl)}\}
     &\binom{1+2}2\binom{2+1}1=9
      & \text{Laurent} \\[1mm] \hline \rule{0pt}{13pt}
 \MR2*{$\geqslant4$}
    & \{\ve_{\9}^{i(j_1\9}\vf_{(a}^{j_2{\cdots}j_{2m-3})}\,
                           p^{\9}_{b)(ij_{m-3}{\cdots}j_{2m-3})}\}
     &\binom{1+2}2(2m{-}3)
      & \text{Laurent} \\
    & \vf_{(a}^{(i_1\cdots\,i_{2m-4})}\,p^{\9}_{b)(i_1\cdots\,i_{2m-4})}=0
     &~~-\binom{2+2}2(m{-}3)\leqslant9+\d^{\sss(2)}_m{}^\ddag &\\[1mm] \hline
\MC4l{\text{\footnotesize$^\ddag$~The ``excess'' number of sections $\d^{\sss(2)}_m = \q^m_3\,(m{-}3)$ is computed in~\eqref{e:nPEP1}--\eqref{e:ndm}.}}
  \end{array}
 \label{e:2dFmK*}
\end{equation}
Similarly, in the $n=3$-dimensional case, we have:
\begin{equation}
 \begin{array}{@{}r|l|l|l@{}}
  \boldsymbol{m} &\boldsymbol{H^0(\cF_m,\cK^*),~~\dim\cF_m=3}
   &\textbf{Number} &\textbf{Sections} \\
 \hline\rule{0pt}{13pt}
  0 & \{\f_{(abc)(ij)}/p_{(a}f_{bc)(ij)}\}
     &\binom{3+3}3\binom{2+1}1-\binom{2+3}3\binom{2+1}1=30
      & \text{ordinary} \\[1mm] \hline \rule{0pt}{12pt}
  1 & \{\f_{(abc)\,i}/f_{(ab}\,p_{c)\,i}\}
     &\binom{3+3}3\binom{1+1}1-\binom{2+3}3\binom{0+1}1=30
      & \text{ordinary} \\[1mm] \hline \rule{0pt}{12pt}
 \MR2*2
    & \{\f_{(abc)\,i}\}
     &\binom{3+3}3\binom{0+1}1=20
      & \text{ordinary} \\
    & \{\ve_{\9}^{ij}f_{(ab}\,p_{c)(ik)}\}
     &\binom{2+3}3\binom{0+1}1=10
      & \text{Laurent} \\[1mm] \hline \rule{0pt}{13pt}
  3 & \{\ve_{\9}^{i(j\9}f_{(ab}^{kl)}\,p^{\9}_{c)(ikl)}\}
     &\binom{2+3}3\binom{2+1}1=30
      & \text{Laurent} \\[1mm] \hline \rule{0pt}{13pt}
 \MR2*{$\geqslant4$}
    & \{\ve_{\9}^{i(j_1\9}\vf_{(ab}^{j_2{\cdots}j_{2m-3})}\,
                           p^{\9}_{c)(ij_{m-3}{\cdots}j_{2m-3})}\}
     &\binom{2+3}3(2m{-}3)
      & \text{Laurent} \\
    & \vf_{(ab}^{(i_1\cdots\,i_{2m-4})}\,p^{\9}_{c)(i_1\cdots\,i_{2m-4})}=0
     &~\quad-\binom{3+3}3(m{-}3)\leqslant30+\d^{\sss(3)}_m{}^\ddag &\\[1mm] \hline
\MC4l{\text{\footnotesize$^\ddag$~The ``excess'' number of sections $\d^{\sss(3)}_m = \q^m_3\,4(m{-}3)$ is computed in~\eqref{e:nPEP1}--\eqref{e:ndm}.}}
  \end{array}
 \label{e:3dFmK*}
\end{equation}

\paragraph{The ``excess'' number of anticanonical sections:}
For completeness, the $m$-twisted Hirzebruch $n$-fold $\sF^{\sss(n)}_m$ may be identified with the projectivization $\sF^{\sss(n)}_m=P(E)$ of the rank-$n$ bundle $E=\cO_{\IP^1}\oplus\cO_{\IP^1}(m)^{\oplus(n-1)}$. The push-forward (to the base-$\IP^1$, where all vector bundles decompose as a direct sum of line-bundles) of the anticanonical bundle of $\sF^{\sss(n)}_m$ is then computed\footnote{We thank Richard Wentworth for alerting us to this independent and standard algebro-geometric computation for Hirzebruch 2-folds, which we generalize here for all Hirzebruch $n$-folds.} as
\begin{subequations}
 \label{e:nPEP1}
\begin{alignat}9
  \pi_*(\cK^*_{\smash{\sF^{\sss(n)}_m}})
   &=(E^*)^n \otimes \big(\cK^*_{\IP^1} \otimes \det(E)\big),\qquad
      E=\cO_{\IP^1}\oplus\cO_{\IP^1}(m)^{\oplus(n-1)};\\
   &=\Big(\bigoplus_{k=0}^n{\ttt\binom{n+k-2}{k}}\cO_{\IP^1}(-km)\Big)
      \otimes \Big(\cO_{\IP^1}(2) \otimes \cO_{\IP^1}\big((n{-}1)m\big)\Big),\\
   &=\bigoplus_{k=0}^n{\ttt\binom{n+k-2}{k}}\cO_{\IP^1}\big(2{+}(n{-}k{-}1)m\big).
\end{alignat}
\end{subequations}
This produces the number of sections, which when combined with the long exact cohomology sequence~\eqref{e:nmH*Q} guarantees that:
\begin{equation}
 \begin{aligned}
  \dim H^0(\sF^{\sss(n)}_m,\cK^*)
   &=\sum_{k=0}^n{\ttt\q_{(k+1)m}^{3+nm}\binom{n+k-2}{k}\,\big(3{+}(n{-}k{-}1)m\big)}
     +\d^{\sss(n)}_m,\\
  \dim H^1(\sF^{\sss(n)}_m,\cK^*)
   &=\d^{\sss(n)}_m := \q_3^m\,{\ttt\binom{2n-2}{n}\,\big(m{-}3\big)},
 \end{aligned}
 \label{e:ndm}
\end{equation}
and $\dim H^i(\sF^{\sss(n)}_m,\cK^*)=0$ for $i>1$. Note that
\begin{equation}
  \chi(\cK^*)=\sum_{i=0}^n \dim H^i(\sF^{\sss(n)}_m,\cK^*)
   = 9,~~30,~~105\quad\text{for}\quad n=2,3,4.
\end{equation}

\subsection{Discrete deformations}
\label{s:DDa}
In fact, not only is it known that the Hirzebruch surfaces $\fF_m$ and $\fF_{m+2}$ are abstractly diffeo\-morphic, one can construct an explicit deformation family of Hirzebruch surfaces that includes both $\fF_0=\IP^1\,{\times}\,\IP^1$ and $\fF_2$~\cite{rGHSAR} and~\cite[Section~3.1.2]{rBeast}.
 This construction provides a complex 1-parameter family such that $\fF_0$ is fibered over $\e\neq0$, while $\fF_2$ fits at $\e=0$: The deformation family is the configuration
\begin{equation}
  \K[{c||cc}{\IP^3&1&1\\\IP^1&1&1\\}]:\qquad
  \BM{x^0 & x^1\\ x^2 & \big(\sum_{i=0}^2a_ix^i{+}\e x^3\big)\\}
   \BM{y^0\\y^1\\} = \BM{0\\0\\}.
 \label{e:P3|11}
\end{equation}
For generic choices of $(a_0,a_1,a_2)$ and $\e\neq0$, the determinant of the system is a smooth quadric in $\IP^3$, known to be the Segr\'e embedding of $\fF_0=\IP^1\,{\times}\,\IP^1=[\IP^3||2]$. At $\e=0$, the determinant of the system develops a singularity, which is ``blown-up'' in the smooth 2-fold $\fF_2$ defined in the $\e\to0$ limit of~\eqref{e:P3|11}.

The deformation space of~\eqref{e:P3|11} is explicitly parametrized by $(a_0,a_1,a_2,\e)\in\IC^4$ and reduces through $\IP^3$-reparametrizations to two distinct but infinitesimally close points: $\{\e\,{\neq}\,0\}$ and $\{\e\,{=}\,0\}$. By this explicit construction, $\lim_{\e\to0}\fF_0=\fF_2$ is a discrete deformation.\footnote{Ref.~\cite{rBeast} calls this a ``jumping deformation.''} Whereas $\fF_0$ and $\fF_2$ are diffeomorphic to each other, we note that there do exist subtle differences:
 $\fF_2$ has an exceptional curve of self-intersection $-2$ and so may be blown down, while $\fF_0=\IP^1\,{\times}\,\IP^1$ does not and so cannot be blown down.
 Owing to the diffeomorphism $\fF_2\approx\fF_0$, we do not expect such a subtle difference to be detectable by any classical field theory model using these spaces, but conjecture that quantum field theory {\em\/may\/} depend on such a subtle difference.

We are not aware of any explicit demonstration that $\sF^{\sss(n)}_m$ and $\sF^{\sss(n)}_{m+n}$ are also discrete deformations of each other for $n\neq2$. However, explicit computation using
\begin{equation}
 \begin{array}{cccccl}
 && T_A\otimes\cP^*_A &\too{\rd p} & \cO_A &
  \MR4*{$\left.\rule{0pt}{30pt}\right\}$ $\sF^{\sss(n)}_m$-resolution}\\
 &&\WC{\into}\raisebox{2pt}{$\SSS\,p$} &&\WC{\into}\raisebox{2pt}{$\SSS\,p$} \\[-1mm]
 &&T_A &\too{\rd p} & \cP_A\\
 &&\WC{\onto}\raisebox{2pt}{$\SSS\rho$} &&\WC{\onto}\raisebox{2pt}{$\SSS\rho$} \\[-2mm]
  T_{\sF^{\sss(n)}_m} &\into &T_A|_{\sF^{\sss(n)}_m} &\overset{\rd p}\onto &\cP|_{\sF^{\sss(n)}_m} &~\quad\text{:~adjunction}
 \end{array}
 \label{e:KrAdjFnm}
\end{equation}
produces:
\begin{equation}
   \dim H^0(\sF^{\sss(n)}_m,T)=n^2{+}2 +\Delta^{\sss(n)}_m \quad\text{and}\quad
   \dim H^1(\sF^{\sss(n)}_m,T)=\Delta^{\sss(n)}_m.
\end{equation}
Here $\Delta^{\sss(n)}_m$ is the net number of Kodaira-Spencer deformations~\cite{rK-DefT} of $\sF^{\sss(n)}_m$ represented, by the tensor components $\f_{a\,(i_1\cdots\,i_m)}$ that cannot be gauged away by the combined transformation
\begin{equation}
 \begin{aligned}
 \d\,\f_{a\,(i_1\cdots\,i_m)}
  &= \l\,p_{a\,(i_1\cdots\,i_m)}
    +\l_a{}^b\,p_{b\,(i_1\cdots\,i_m)}
    + p_{a\,j(i_1\cdots\,i_{m-1}}\,\l_{i_m)}{}^j\\
  &\quad + \q_2^m\,p_{a\,j(i_1\cdots\,i_{m-1}|}\,p_{b\,|i_m)(k_1\cdots\,k_{m-1})}\,
    \l^{b\,(k_1\cdots\,k_{m-2}}\ve^{k_{m-1})j}.
 \end{aligned}
 \label{e:gtr}
\end{equation}
Contracting this equivalence class of tensors with $x^a\,y^{i_1}\,{\cdots}\,y^{i_m}$ provides this tensorial relation with the familiar interpretation of reparametrizations of the polynomial deformations --- except for the higher cohomology contribution in the last term containing the step-function $\q_2^m$.

 In turn, the tensor components
 $\{\l,~\l_a{}^b,~\l_i{}^j,~\l^{a\,(i_1\cdots\,i_{m-2})}:~\l_a{}^a=0=\l_i{}^i\}$
that cannot be used up in the transformation~\eqref{e:gtr} span $H^0(\sF^{\sss(n)}_m,T)$ --- representing the coordinate reparametrizations of $\sF^{\sss(n)}_m$. Exceptionally for $m=0$, the $\IP^n$ reparametrization generators $\l_a{}^b$ are themselves subject to an additional equivalence relation generated by multiplication by the defining tensor $p_a$, and in addition to the constraints~\eqref{e:gtr}. It is gratifying to note that the dual constraint sub-system for $m=0$:
\begin{equation}
   \{\l_a{}^b\simeq \l_a{}^b+p_a\,\vq^b\}:~\l_a{}^b\,p_b=0
\end{equation}
leaves $\{[(n{+}1)^2{-}1]-[n{+}1]\}-n = n^2{-}1$ free components of $\l_a{}^b$, as is appropriate for $\IP^{n-1}$ coordinate reparametrizations, in
 $\ssK[{l||c}{\IP^n&1\\ \IP^1&0}]=\IP^{n-1}\,{\times}\,\IP^1$. The $\vq^a$-generated equivalence allows gauging away $n{+}1$ degrees of freedom, but the constraint system $\l_a{}^b\,p_b=0$ consists of only $n$ independent equations, since $\l_a{}^b\,p_b$ cannot be proportional to $p_a$ itself as $\l_a{}^a=0$ and so $\l_a{}^b\not\propto\d_a^b$.

It is known~\cite{rBeast} that for $\fF_0$, $\fF_1$ and $\fF_2$ are rigid but $\Delta^{\sss(n)}_m$ may well be nonzero for other values of $m$ and $n$. For the general case $m\geqslant0$ and $n\geqslant2$, the analysis of the constrained gauge-equivalence system of tensors~\eqref{e:gtr} becomes considerably more involved, and we leave the precise determination of $\dim H^1(\sF^{\sss(n)}_m,T)=\Delta^{\sss(n)}_m$ --- and therefore also of $\dim H^0(\sF^{\sss(n)}_m,T)=n^2{+}2 +\Delta^{\sss(n)}_m$ for a subsequent effort.

Therefore, we have that for all $m\geqslant0$ and for $n=2,3$, the Hirzebruch $n$-folds $\sF^{\sss(n)}_m\approx\sF^{\sss(n)}_{m+n}$ are diffeomorphic; we {\em\/assume\/} this to be true also for $n=4$ at least.
Given that $\fF_m$ and $\fF_{m+2}$ in fact are discrete deformations of each other, we propose:
\begin{conj}\label{c:DDeFm}
({\bfseries i})~The $\Delta^{\sss(n)}_m$-dimensional deformation spaces of ``adjacent'' Hirzebruch $n$-folds
 $\sF^{\sss(n)}_m=\ssK[{c||c}{\IP^n&1\\\IP^1&m}]$ and
 $\sF^{\sss(n)}_{m+n}=\ssK[{c||c}{\IP^n&1\\\IP^1&m{+}n}]$
to be ``separate but infinitesimally near,'' so that $\sF^{\sss(n)}_m$ is a discrete deformation of $\sF^{\sss(n)}_{m+n}$.

({\bfseries ii})~In any classical field theory, the use of $\sF^{\sss(n)}_m$ and $\sF^{\sss(n)}_{m+n}$ should produce identical models; however, some quantum effects may well distinguish $\sF^{\sss(n)}_m$ from $\sF^{\sss(n)}_{m+n}$.
\end{conj}

\section{The refined Koszul resolution for meromorphic intersections}
\label{s:gTESS}
The Koszul resolution~\eqref{e:KXm} of the holomorphic sheaf of functions over $X_m$ may be written as
\begin{equation}
  \cP^*\,{\otimes}\,\cQ^*
   \vC{\begin{picture}(27,14)(0,-7)
         \put(2,.5){\vector(2,1){8}}  \put(5,4){$\SSS p$}
         \put(2,-.5){\vector(2,-1){8}} \put(5,-5){$\SSS q$}
         \put(12,4){$\cQ^*$}
         \put(13,-1){$\doo{\SSS\ve\!f}$}
         \put(12,-6){$\cP^*$}
         \put(18,4.5){\vector(2,-1){8}} \put(21,4){$\SSS q$}
         \put(18,-4.5){\vector(2,1){8}} \put(21,-5){$\SSS p$}
       \end{picture}}
   \cO_A \overset{\rho}\onto\cO_{X_m}.
 \label{e:KXm}
\end{equation}
where the $p$- and the $q$-maps are contractions with the defining tensors $p_{a\,(i_1\cdots\,i_m)}$ from~\eqref{e:fDef} and $q_{(abcd)\cdots}^{\cdots}$~\eqref{e:q=efp}. Generalizing~\eqref{e:KFm}, this identifies holomorphic objects $\cB$ on $X_m$ as the analogous objects $\cB$ defined on $A=\IP^4\,{\times}\,\IP^1$, taken however modulo $p$-multiples of $\cB\otimes\cP^*$ and $q$-multiples of $\cB\otimes\cQ^*$, and taking into account the ``double-counting'' of objects that are $p{\cdot}q$-multiples of $\cB\otimes\cP^*\otimes\cQ^*$.

However, we must also include the {\em\/additional\/} bundle map, $\too{\ve\!f}$, induced by the contraction~\eqref{e:2q=efp} and~\eqref{e:3q=efp}: linear maps generated by contracting with the $q$-tensor~\eqref{e:q=efp} equal the sequential contraction with $p_{a(i_1\cdots\,i_m)}$- and $\ve_{\9}^{ij}f^{(k_1\cdots\,k_{2m-4})}_{(abc)}$-tensors, in either order. However, since the corresponding cohomology $\ve\!f$-map involves contracting with (the $\rd$-preimage of) an element of $H^1(A,\cQ\,{\times}\,\cP^*)$, it can be nonzero only for $m\geqslant2$ when $H^1(A,\cQ\,{\times}\,\cP^*)\neq0$, and must act $H^r(A,\cB)\too{\ve\!f}H^{r+1}(A,\cB\,{\otimes}\,\cQ\,{\otimes}\,\cP^*)$.
 In this way, we extend the standard treatment of algebraic systems of constraints~\cite{rBeast} to gCICYs.

We then tensor~\eqref{e:KXm} by each of the components of the tangent and normal bundles,
\begin{subequations}
 \label{e:TQn}
\begin{alignat}9
   T_{\IP^4\oplus\IP^1} &= T_{\IP^4} \oplus T_{\IP^1},\\
   N &= (\cP=\cO\pM{1\\m\\}) \oplus (\cQ=\cO\pM{4\\\!2-m\!\\}),
\end{alignat}
\end{subequations}
to compute the so-valued cohomology on $X_m$, for use in the cohomology sequence associated with the adjunction monad~\eqref{e:Adjunct}.

\subsection{The $T$-valued cohomology}
\label{s:T}
Being formed from sections of $\cQ=\cO\pM{4\\\!2-m\!\\}$, $q(x,y)$ is non-positive over $\IP^1$ for $m>2$, and we cannot use the Lefschetz hyperplane theorem~\cite{rB+S-Adjunction,rBeast} to compute the Hodge numbers $h^{*,*}(X_m)$. However, we can compute $H^*(X_m,T)=H^{2,*}(X_m)$ and $H^*(X_m,T^*)=H^{1,*}(X_m)$ using respectively the ``adjunction'' sequence and its dual:
\begin{equation}
  T_{X_m} \into T_A|_{X_m} \overset{\rd q}\onto [\cP\,{\oplus}\,\cQ]_{X_m}
    \qquad\text{and}\qquad
  [\cP^*\,{\oplus}\,\cQ^*]_{X_m} \overset{\rd q}\into T^*_A|_{X_m} \onto T^*_{X_m}.
 \label{e:Adjunct}
\end{equation}
The restrictions $\cB|_{X_m}=\rho(\cB)$ are obtained using the codimension-2 Koszul resolution~\eqref{e:KXm} of $X_m\subset A$. In particular, the Koszul resolutions of $\cP=\cO\pM{1\\m\\}$ is (crossed-out sheaves have no cohomology):
\begin{equation}
\begin{array}{c|ccc@{~}|@{~}c}
   & ~~\str{14}\cO\pM{-4\\\!m-2\!\\}~~
    & \MM{ \str{16}\cO\pM{-3\\\!2m-2\!\\}\\[1mm]
            \cO\pM{0\\0\\}\piC{\put(1,1){\vector(3,1){8}}
                               \put(5,0){$\SSS p$}}\\}
      & \cO\pM{1\\m\\} & \cP\big|_{X_m} \\[4mm]
 \hline\rule{0pt}{3ex}
 0. & 0 & \{\vf\}\piC{\put(1,0){$\too{\,p\,}$}}
           & \{\f_{a(i_1\cdots\,i_m)}\} & H^0(X_m,\cP) \\ 
 1. & 0 & 0 & 0 & 0 \\ 
 \vdots & \vdots & \vdots & \vdots & \vdots \\ 
 \hline
\end{array}
 \quad
 \begin{array}{l}
  H^0(X_m,\cP)\\ \sim
  \{\f_{a(i_1\cdots\,i_m)}/p_{a(i_1\cdots\,i_m)}\,\vf\}\\[1mm]
  \dim=\binom{1+4}4{\cdot}\binom{m+1}1-1\\
  \hphantom{\dim}=5(m{+}1)-1
 \end{array}
 \label{e:P|Xm}
\end{equation}
The Koszul resolution of $\cQ=\cO\pM{4\\\!2-m\!\\}$ is more involved:
\begin{equation}
\begin{array}{c|ccc@{~}|@{~}c}
   & ~~\str{14}\cO\pM{\!-1\\\!-m\\}~~
    & \MM{ \cO\pM{0\\0\\}\\[-.5mm]
            \doo{\ve\!f} \\[-.5mm]
             \cO\pM{3\\\!2-2m\!\\} \\}
              \piC{\put(-2,5){\rotatebox{-7.5}{\vector(1,0){28}}}
                    \put(12,4){$\SSS q$}
                   \put(1,-3){\rotatebox{7.5}{\vector(1,0){24.5}}}
                    \put(12,-3){$\SSS p$} }\quad{~~~}
      & \cO\pM{4\\\!2-m\!\\} & \cQ\big|_{X_m} \\[6mm]
  \hline\rule{0pt}{5ex}
 0. & 0 & \MM{\q_m^1\{\vf_{(abc)(i_1\cdots i_{2-2m})}\}
              \piC{\put(1,1){\vector(4,-1){7}}\put(5,1){$\SSS p$}}\\[1mm]
             \{\vq\}
               \piC{\put(1,.5){\rotatebox{10}{\vector(1,0){20}}}
                    \put(13,1){$\SSS q$}} \\}
           &~\quad~\q_m^2\{\f_{(abcd)(i_1\cdots\,i_{2-m})}\} & H^0(X_m,\cQ) \\[-.5mm] 
 && \doo{\ve\!f} && \\[-.5mm]
 1. & 0 & \q^m_2\{\ve^{ij}\vf_{(abc)}^{(k_1\cdots\,k_{2m-4})}\}
          \piC{\put(1,1){\vector(1,0){9}} \put(5,2){$\SSS p$}
                \blue{\mdot(-8,4){11}{32}} }
           &\quad\q^m_4\{\g_{(abcd)}^{(i_1\cdots\,i_{m-2})}\} & H^1(X_m,\cQ) \\ 
 2. & 0 & 0 & 0 & 0 \\ 
 \vdots & \vdots & \vdots & \vdots & \vdots \\ 
 \hline
\end{array}
 \label{e:Q|Xm}
\end{equation}
The $H^0(X,\cQ)$ cohomology group varies with $m$ and is represented by tensors in the following $m$-dependent fashion:
\begin{equation}
 \begin{array}{c|l}
\boldsymbol{m} & \textbf{Tensor Representative of }\boldsymbol{H^0(X,\cQ)} \\
  \hline\rule{0pt}{2.5ex}
0 & \big\{\f_{(abcd)(ij)}\big/
           \big[p_{(a}\,\vf_{bcd)(ij)} \oplus \vq\,q_{(abcd)(ij)}\big]\big\} \\[1mm]
1 & \big\{\f_{(abcd)\,i}\big/
           \big[\vf_{(abc}\,p_{d)\,i} \oplus \vq\,q_{(abcd)\,i}\big]\big\} \\[1mm]
2 & \big\{\big[\f_{(abcd)}\oplus\ve^{ij}\vf_{(abc}\,p_{d)ik}\big]\big/
            \vq\big[q_{(abcd)}\oplus\ve^{ij}f_{(abc}\,p_{d)ik}\big]\big\} \\[1mm]
3 & \big\{\big[\ve^{i(j}_{\9}\vf^{kl)}_{(abc}\,\big/\,
            \vq\,\ve^{i(j}_{\9}f^{kl)}_{(abc}\big]\,p^{\9}_{d)(jkl)}\big\} \\[1mm]
\geqslant4
  & \Big\{\big[\ve^{i(j_1}_{\9}\f^{j_2\cdots j_{2m-3})}_{(abc}\big/
            \vf\,\ve^{i(j_1}_{\9}f^{j_2\cdots j_{2m-3})}_{(abc}\big]\,
             p^{\9}_{d)(ij_{m-1}\cdots j_{2m-3})},\\
  &\quad \text{s.t.}: f^{(j_1{\cdots}j_{2m-4})}_{(abc}\,
                       p^{\9}_{d)(j_{m-3}{\cdots}j_{2m-4})}=0\Big\} \\[1mm]
 \end{array}
 \label{e:mH0Q}
\end{equation}

Denoting for brevity $T_i\pM{d_1\\[1pt]\smash{d_2}\\[1pt]}\define T_i\,{\otimes}\,\cO\pM{d_1\\[1pt]\smash{d_2}\\[1pt]}$, we tensor~\eqref{e:KXm} by $T_{\IP^4}$:
\begin{equation}
\begin{array}{c|ccc@{~}|@{~}c}
   & T_{\IP^4}\pM{\!-5\\\!-2\\}
    & \MM{ \str{18}T_{\IP^4}\pM{-4\\\!m-2\!\\}\\[1mm]
           T_{\IP^4}\pM{\!-1\\\!-m\\}\piC{ \put(1,1){\vector(4,1){13}}
                                       \put(6,3.5){$\SSS p$} }\\}
      & T_{\IP^4} & T_{\IP^4}\big|_{X_m} \\[4mm]
 \hline\rule{0pt}{3ex}
 0. & 0 & \d_{m,0}\,\{\vk^a\}\piC{ \put(1,0){$\too{~~~p~~~}$} }
           &~~~\{\l_a{}^b\}~~~ & H^0(X_m,T_{\IP^4}) \\[1mm]
 1. & 0 & \q^m_2\{\ve^{i(j_1}\vk^{j_2\cdots\,j_{m-1})\,a}\}
          \piC{\blue{\mdot(-3,3){11}{9}}}
           & 0 & 0 \\ 
 2. & 0 & 0 & 0 & H^2(X_m,T_{\IP^4}) \\ 
 3. & 0 & 0 & 0 & 0 \\ 
 4. &\{\ve^{abcde}\ve^{ij}\Lambda_1\}\piC{\blue{\mdot(0,2.5){11}{45}}}
       & 0 & 0 & \text{---} \\ 
 5. & 0 & 0 & 0 & \text{---} \\ 
 \hline
\end{array}
 \label{e:T1|Xm}
\end{equation}
This lets us represent
\begin{alignat}9
 H^0(X_m,T_{\IP^4})&:~~\{\l_a{}^b/\d_{m,0}(p_a\vk^b)\}
    \oplus \q_2^m\{\tilde\vk_{k\9}^{~i}\define
                    \ve^{i(j_1}\vk^{j_2\cdots\,j_{m-1})\,a}p_{a\,(j_1\cdots\,j_{m-1}k)}\},
 \label{e:H0TP4}\\
 H^2(X_m,T_{\IP^4})&:~~\{\ve^{abcde}\ve^{ij}\Lambda_1\}. \label{e:H2TP4}
\end{alignat}
Except for $m=0$, the first contribution to $H^0(X_m,T_{\IP^4})$ represents the standard linear, traceless $\IP^4$-reparametrizations $\{\l_a{}^b\}$. Although the second contribution acts as a standard linear, traceless $\IP^1$-reparametri\-za\-tion, it is parametrized by the $\binom{(m-2)+1}1\binom{1+4}4=5(m{-}1)$-component tensor $\vk^{(j_1\cdots\,j_{m-2})\,a}$.
We note that $\ve^{abcde}\ve^{ij}\Lambda_1$ represents the Serre dual of the K\"ahler form $J_1\in H^2(\IP^4,\ZZ)$.

Tensoring~\eqref{e:KXm} by $T_{\IP^1}$, we obtain:
\begin{equation}
\begin{array}{c|ccc|c}
   & T_{\IP^1}\pM{\!-5\\\!-2\\}
    & \MM{ \str{18}T_{\IP^1}\pM{-4\\m-2\\} \\[1mm]
           \str{16}T_{\IP^1}\pM{\!-1\\\!-m\\}\\}
      & T_{\IP^1} & T_{\IP^1}\big|_{X_m} \\[4mm]
 \hline\rule{0pt}{3ex}
 0. & 0 & 0 &~~~\{\l_i{}^j\}~~~ & H^0(X_m,T_{\IP^1}) \\ 
 1. & 0 & 0 & 0 & 0 \\ 
 2. & 0 & 0 & 0 & H^2(X_m,T_{\IP^1}) \\ 
 3. & 0 & 0 & 0 & 0 \\ 
 4. &\{\ve^{abcde}\ve^{ij}\Lambda_2\}\piC{\blue{\mdot(0,3.25){16}{28}}}
       & 0 & 0 & \text{---} \\ 
 5. & 0 & 0 & 0 & \text{---} \\ 
 \hline
\end{array}
 \label{e:T2|Xm}
\end{equation}
This represents $H^0(X_m,T_{\IP^1})$ by the standard linear, traceless $\IP^1$-reparametrization $\{\l_i{}^j\}$, and $\ve^{abcde}\ve^{ij}\Lambda_2$ represents $H^2(X_m,T_{\IP^1})$ and the Serre dual of the K\"ahler form $J_2\in H^2(\IP^1,\ZZ)$.

Putting~\eqref{e:P|Xm},~\eqref{e:T2|Xm},~\eqref{e:T1|Xm} and~\eqref{e:Q|Xm} together, we obtain:
\begin{equation}
\begin{array}{c|cc}
  T_{X_m} & T_{\IP^4\times\IP^1}\big|_{X_m}
        & [\cP\oplus\cQ]_{X_m} \\[4mm]
 \hline
 H^0(X_m,T) & \MM{ \{\l_a{}^b/\d_{m,0}(p_a\vk^b)\} \\[0mm]
                   \{\l_i{}^j\} \\[0mm]
                   \q_2^m\{\tilde\vk_{k\9}^{~i}\}_{\text{(\ref{e:H0TP4})}} \\ }
       \underset{\rd q}{\overset{\rd p}\to}
              & \begin{array}{l}
            \{\f_{a(i_1\cdots\,i_m)}\,/\vq\,p_{a(i_1\cdots\,i_m)}\}\\[1mm]
            \Bigg\{\underbrace{\begin{array}{l}
               \q_m^2\{\f_{(abcd)(i_1\cdots\,i_{2-m})}\,/\ldots\}\,\oplus\\[1mm]
               \q_2^m\{\g^{(j_1\cdots\,j_{m-2})}_{(abcd)}\,/\ldots\}
            \end{array}}_{\text{see~\eqref{e:4H0} and~\eqref{e:mH0Q}}}\Bigg\}
                \end{array}\\ 
 H^1(X_m,T) & 0 & 0 \\ 
 H^2(X_m,T) & \{\ve^{abcde}\ve^{ij}(\Lambda_1\oplus\Lambda_2)\}& 0 \\ 
 H^3(X_m,T) & 0 & 0 \\ 
 \hline
\end{array}
 \label{e:HqTm}
\end{equation}
The combined mapping $\underset{\rd q}{\overset{\rd p}\to}$ maps the contributions from the $T_{\IP^4\times\IP^1}|_{X_m}$ column to those in the $[\cP\oplus\cQ]_{X_m}$ column by means of a simple contraction with $p_{a(i_1\cdots\,i_m)}$, or $q_{(abcd)(i_1\cdots\,i_{2-m})}$ for $m\leqslant2$, or $q_{(abcd)}^{(i_1\cdots\,i_{m-2})}$ for $m\geqslant3$ and generates the equivalence relations such as:
\begin{subequations}
\begin{alignat}9
 \l_a{}^b&:~~ \f_{a(k_1\cdots\,k_m)}&
  &\simeq\f_{a(k_1\cdots\,k_m)} +\l_a{}^b\,p_{b(k_1\cdots\,k_m)},\\
 \l_i{}^j&:~~ \f_{a(i_1\cdots\,i_m)}&
  &\simeq\f_{a(k_1\cdots\,k_m)} +\l_{k_1}{}^j\,p_{b(jk_2\cdots\,k_m)},\\
 \text{for}~m\geqslant2,~ \tilde\vk_{i\9}^{~j}&:~~ \f_{a(i_1\cdots\,i_m)}&
  &\simeq\f_{a(k_1\cdots\,k_m)} +\tilde\vk_{k_1\9}^{~j}\,p_{b(jk_2\cdots\,k_m)},
\end{alignat}
and similarly with the other terms in $H^0(X_m,\cP\,{\oplus}\,\cQ)$. Notice that the $\tilde\vk^{\,i}{}_j$-reparamet\-ri\-zations acting on the $m\geqslant2$ contributions to $H^0(X_m,\cQ)$ are given as:
\begin{equation}
  \tilde\vk^{\,i}{}_j:~~
   \g^{(j_1\cdots\,j_{m-2})}_{(abcd)}\simeq\g^{(j_1\cdots\,j_{m-2})}_{(abcd)}
   +\tilde\vk_{i\9}^{~j_1\9}\,q^{(ij_2\cdots\,j_{m-2})}_{(abcd)},
 \label{e:vkq}
\end{equation}
\end{subequations}
and involve a contraction with the defining tensor $p_{a\,(i_1\cdots\,i_m)}$ twice: both within the definition~\eqref{e:H0TP4} of $\tilde\vk_{i\9}^{~j_1\9}$ and also within the definition~\eqref{e:q=efp} of $q^{(ij_2\cdots\,j_{m-2})}_{(abcd)}$. Although acting as a standard linear, traceless $\IP^1$-reparametrization, $\tilde\vk_{i\9}^{~j_1\9}\define \ve^{i(j_1}\vk^{j_2\cdots\,j_{m-1})\,a}\,p_{a\,(j_1\cdots\,j_{m-1}k)}$ is parametrized by the $5(m{-}1)$ independent components of $\vk^{(i_1\cdots\,i_{m-2})\,a}$, and can ``gauge away'' that many degrees of freedom via relations such as~\eqref{e:vkq}.

For generic choices of $p(x,y)$ and $q(x,y)$, defined respectively in~\eqref{e:fDef} and~\eqref{e:q=efp}, the combined mapping $\xrightarrow[\rd q]{\rd p}$ in~\eqref{e:HqTm} is of maximum rank, i.e., has no kernel so that $H^0(X_m,T)=0$ and the tangent bundle of $X_m$ is 
simple, as expected for Calabi-Yau $n$-folds on general grounds. In turn, that implies that
\begin{equation}
  H^1(X_m,T) \approx H^0(X_m,\cP\,{\oplus}\,\cQ) \big/ \{\mM{\rd p\\\rd q\\}\}{\cdot}H^0(X_m,T_{\IP^4\times\IP^1}),
\end{equation}
so that $h^{2,1}=\dim H^1(X_m,T)$ is given, for various $m$ as:
\begin{alignat}9
  h^{2,1}
  &=\big\{5(m{+}1)-1\big\} +\big\{\q_m^2[70(3{-}m)-\q_m^1\,35(3{-}2m)]
    +\q_2^m[35(2m{-}3)-\q_4^m\,70(m{-}3)]-1\big\}\nonumber\\
  &~~~~-\big\{(24-\d_{m,0}5)+(3)+\q_2^m[5(m{-}1)]\big\},
\end{alignat}
which evaluates to $h^{2,1}=86$ for all $m\geqslant0$.

Finally,~\eqref{e:HqTm} also shows that $H^2(X_m,T)\approx H^2(X_m,T_{\IP^4\times\IP^1})$ which in turn was shown in~\eqref{e:T1|Xm} and~\eqref{e:T2|Xm} to be generated by the duals of the K\"ahler forms of $\IP^4$ and $\IP^1$. Therefore, $H^{1,1}(X_m,T^*)\approx H^2(X_m,\ZZ)$ is generated by the direct images of those K\"ahler forms.

While rather involved, the above computation is considerably swifter than the monad-by-monad calculation following through the system of eleven monads in Figure~\ref{f:nAdX}. We have however verified that these two computational frameworks do agree, and in particular that the $\too{\ve\!f}$-map modification of the Koszul resolution~\eqref{e:KXm} is both necessary and sufficient in all the cases considered herein.
\begin{figure}[ht]
\begin{equation*}
\begin{array}{ccccccccc}
 \cQ^* &&\Cx{$T_A{\otimes}\cQ^*{\otimes}\cP^*$} \\[-1mm]
 \WC{\into}\raisebox{2pt}{$\SSS\,p$} &&\WC{\into}\raisebox{2pt}{$\SSS\,p$} \\[-1mm]
 \cP{\otimes}\cQ^* &&T_A{\otimes}\cQ^* &&&&T_A{\otimes}\cP^* &&\cO_A \\[-1mm]
 \WC{\onto} &&\WC{\onto} &&&&\WC{\into}\raisebox{2pt}{$\SSS\,p$}
                            &&\WC{\into}\raisebox{2pt}{$\SSS\,p$} \\[-2mm]
 [\cP{\otimes}\cQ^*]_{F_m} &\overset{\rd p}{\BW{\onto}} &[T_A{\otimes}\cQ^*]_{F_m}
                        &\BW{\into}&T_{F_m}{\otimes}\cQ^*_{F_m} &&T_A &&\cP \\[-1mm]
 &&&&\WC{\into}\raisebox{2pt}{$\SSS\,q$} &&\WC{\onto} &&\WC{\onto} \\[-2mm]
 &&&&T_{F_m} &\into&T_A|_{F_m} &\overset{\rd p}{\onto} &\cP_{F_m} \\[-1mm]
 &&&&\WC{\onto} &&&& \\[-2mm]
 &&\fbox{$T_{X_m}$}&\into&T_{F_m}|_{X_m} &\overset{\rd q}{\onto}&\cQ_{F_m}|_{X_m}&& \\[-1mm]
 &&&&&&\CW{\onto}&& \\[-1mm]
 &&\cP^*{\otimes}\cQ&\overset{p}{\into}&\cQ&\onto&\cQ_{F_m}&& \\[0mm]
 &&&&&&\CW{\into}{\SSS q}&& \\[-1mm]
 &&\cP^*&\overset{p}{\into}&\cO_A&\onto&\cO_{F_m}&& \\[-2mm]
\end{array}
\end{equation*}
\caption{The network of eleven monads ($A\into B\onto C$, i.e., $C=B/A$) that determine the tangent bundle $T_{X_m}$ on $X_m\subset\IP^4\,{\times}\,\IP^1$ in terms of various bundles and sheaves on $A=\IP^4\,{\times}\,\IP^1$.}
 \label{f:nAdX}
\end{figure}

\section{Further examples}
\label{s:More}
\subsection{A terminating sequence}
\label{s:noSeq}
To demonstrate the relevance of checking the number and tensorial structure of anticanonical sections as done in~\eqref{e:4dFmK*}, as well as in~\eqref{e:2dFmK*} and~\eqref{e:3dFmK*}, let us consider the common zero-locus of the system of equations $p(x,y)=0=q(x,y)$:
\begin{equation}
  \Tw{X}_m \in
  \K{[}{c||c|c}{
   \IP^4 & 2 & 3 \\
   \IP^1 & m & 2{-}m \\}{]}^{(2,56-15m)}_{-108 + 30 m},
  \qquad m=0,1,2,3,\ldots;\qquad
  \Tw{X}_m\subset \Tw{F}_m \in
  \K{[}{c||c}{
   \IP^4 & 2 \\
   \IP^1 & m \\}{]}.
 \label{e:SeqT}
\end{equation}
For $m>2$, the second, degree-$\pM{3\\\!2-m\!\\}$ equation is negative over $\IP^1$ and we need to verify that $\cQ=\cO_{\Tw{F}_m}\pM{3\\\!2-m\!\\}$ has sections. As in Appendix~\ref{s:-K}, the Koszul resolution for $\Tw{F}_m$ provides the required restriction:
\begin{equation}
  \begin{array}{c|ccccl}
   & \cO\pM{1\\\!2-2m\!\\} &\overset{p}\into 
   & \cQ=\cO\pM{3\\\!2-m\!\\} &\overset{\rho_G}\onto
   & \cQ|_{\Tw{F}_m}\\*[1mm] \hline \rule{0pt}{6mm}
 0. &\q_m^1\{\vf_{a\,(i_1\cdots\,i_{2-2m})}\} &\too{p}
      &\q_m^2\{\f_{(abc)(i_1\cdots\,i_{2-m})}\} &\too{\rho_G}
          & H^0(\Tw{F}_m,\cQ)\too{\rd} \\*[2mm]
 1 &\q_2^m\{\ve_{\9}^{i(j}\vf_a^{k_1\cdots\,k_{2m-4})}\} &\too{p}
      &\q_4^m\{\ve_{\9}^{i(j}\f_{(abc)}^{k_1\cdots\,k_{m-4})}\} &\too{\rho_G}
          & H^1(\Tw{F}_m,\cQ)\too{\rd} \\*
 2 & 0 && 0 && H^2(\Tw{F}_m,\cQ)=0 \\*[-2mm]
 \vdots & \vdots && \vdots &&~~~ \vdots \\ 
    \hline
 \MC6l{\text{\small The $p$-map is generated by contraction with $p_{(ab)(i_1\cdots\,i_m)}$}}
  \end{array}
\end{equation}
A section $q(x,y)\in H^0(\Tw{F}_m,\cQ)$ is to be used for the defining equation of $\Tw{X}_m\subset\Tw{F}_m$. For $m\leqslant3$, $H^0(\Tw{F}_m,\cQ)$ is nonzero:
\begin{equation}
 \begin{array}{@{}r|l|l|l@{}}
  \boldsymbol{m} &\boldsymbol{H^0(\Tw{F}_m,\cQ)} &\textbf{Number} &\textbf{Sections} \\
 \hline\rule{0pt}{13pt}
  0 & \{\f_{(abc)(ij)}/p_{(ab}\vf_{c)}(ij)\}
     &\binom{3+4}4\binom{2+1}1-\binom{1+4}4\binom{2+1}1=90
      & \text{ordinary} \\[1mm] \hline \rule{0pt}{12pt}
  1 & \{\f_{(abc)\,i}/\vf_{(a}\,p_{bc)\,i}\}
     &\binom{3+4}4\binom{1+1}1-\binom{1+4}4\binom{0+1}1=65
      & \text{ordinary} \\[1mm] \hline \rule{0pt}{12pt}
 \MR2*2
    & \{\f_{(abc)}\}
     &\binom{3+4}4\binom{0+1}1=35
      & \text{ordinary} \\
    & \{\ve_{\9}^{ij}\vf_{(a}\,p_{bc)(ik)}\}
     &\binom{1+4}4\binom{0+1}1=5
      & \text{Laurent} \\[1mm] \hline \rule{0pt}{13pt}
  3 & \{\ve_{\9}^{i(j\9}\vf_{(a}^{kl)}\,p^{\9}_{bc)(ikl)}\}
     &\binom{1+4}4\binom{2+1}1=15
      & \text{Laurent} \\ \hline
  \end{array}
\end{equation}
However, for $m\geqslant4$ the cohomology group $H^0(\Tw{F}_m,\cQ)$ vanishes for maximal-rank choices of the tensor coefficients in the defining sections $p(x,y)$ and $q(x,y)$, so that
\begin{equation}
 \begin{aligned}
  H^1(\Tw{F}_m,\cQ)
   &\sim\{\ve_{\9}^{i(j}\f_{(abc)}^{k_1\cdots\,k_{m-4})}/
           \ve_{\9}^{i(j}\vf_a^{k_1\cdots\,k_{2m-4})}\},\\
  \dim H^1(\Tw{F}_m,\cQ)
   &=\ttt\binom{3+4}4\binom{m-4+1}1-\binom{1+4}4\binom{2m-4+1}1=5(13m-18).
 \end{aligned}
\end{equation}
Unlike~\eqref{e:Seq}, the sequence of Calabi-Yau 3-folds~\eqref{e:SeqT} terminates with $\Tw{X}_3$: for $m\geqslant4$, $\ssK{[}{c||c}{\IP^4&2\\ \IP^1&m\\}{]}$ has no holomorphic sections of $\cQ=\cO\pM{3\\\!2-m\!\\}$ with which to define $\Tw{X}_m$.
 In turn, the computation analogous to the one presented in Appendix~\ref{s:T} insures that $H^{1,1}(\Tw{X}_m)$ is 2-dimensional and is generated by (the pullbacks of) the K\"ahler classes of $\IP^4$ and $\IP^1$ for all $m\geqslant2$. This justifies the standard Chern-class computation and produces $\chi_{_E}$, $h^{1,1}$ and so also $h^{2,1}$ as displayed in~\eqref{e:SeqT}.

In particular, $h^{2,1}$ would become (nonsensically) negative for $m\geqslant4$ --- when in fact there are no holomorphic sections to define $\Tw{X}_m\subset \Tw{F}_m$ in the first place. In this way, the termination of the series is signaled already by the standard Chern-class computation.

\subsection{More complicated configurations}
\label{s:More2}
Even just permitting the Hirzebruch $n$-folds of arbitrary twist as a factor in the embedding space generalizes the constructions of both Refs.~\cite{rH-CY0,rGHCYCI,rCYCI1} as well as Ref.~\cite{rGHSAR} with doubly periodic examples such as:
\begin{subequations}
 \label{e:dblP}
\begin{alignat}9
 \text{$[m\pMod2]{\cdot}[n\pMod2]$:}&\quad
  \K[{c||cc|c}
  {\IP^2 & 1 & 0 &\piC{\put(1,1){\color{blue}\circle{5}}}2  \\
   \IP^1 & m & 0 & 2{-}m\\[1pt]
    \hdashline[1pt/1pt]\rule{0pt}{12pt}
   \IP^2 & 0 & 1 & \piC{\put(1,1){\color{blue}\circle{5}}}2  \\
   \IP^1 & 0 & n & 2{-}n\\}]_{-128}^{(4,68)},
 \label{e:dblPa}\\
 \text{$[m\pMod2]{\cdot}[n\pMod3]$:}&\quad
  \K[{c||cc@{~~}c|c}
  {\IP^2 & 1 & 0 & 0 & \piC{\put(1,1){\color{blue}\circle{5}}}2  \\
   \IP^1 & m & 0 & 1 & 1{-}m\\[1pt]
    \hdashline[1pt/1pt]\rule{0pt}{12pt}
   \IP^3 & 0 & 1 & 0 & \piC{\put(1,1){\color{blue}\circle{5}}}3  \\
   \IP^1 & 0 & n & 1 & 1{-}n\\}]_{-144}^{(4,76)},
 \label{e:dblPb}
\end{alignat}
\end{subequations}
which are both doubly periodic. By direct computation as in Appendices~\ref{s:HFm} and~\ref{s:gTESS}, we verify that the last constraint is viable, i.e., that the regular complete intersection 4-fold defined by imposing all but this last constraint has sections of the degrees specified in the last column:~\eqref{e:dblPa} has 81, while~\eqref{e:dblPb} has 90 sections with which to define this last constraint and so the indicated Calabi-Yau 3-folds.

Among several different ways to regard these configurations and as indicated by the horizontal dashed lines, the configuration~\eqref{e:dblPa} may be regarded as a fibration of the ``upper'' torus $\ssK[{c||c|c}{\IP^2&1&2\\\IP^1&m&2{-}m\\}]$ over the ``lower'' $\fF_n$ or as a fibration of the ``lower'' torus $\ssK[{c||c|c}{\IP^2&1&2\\\IP^1&n&2{-}n\\}]$ over the ``upper'' $\fF_m$. Similarly, the configuration~\eqref{e:dblPb} may be regarded as a fibration of the ``upper'' 2-points $\ssK[{c||c|cc}{\IP^2&1&0&2\\\IP^1&m&1&2{-}m\\}]$ over the ``lower'' $\cF_n$ or as a fibration of the ``lower'' torus $\ssK[{c||c|cc}{\IP^3&1&0&3\\\IP^1&n&1&2{-}n\\}]$ over the ``upper'' $\fF_m$.

To demonstrate that the particular degrees indicated by circles in~\eqref{e:dblP} are necessary for the Hirzebruch $n$-folds' periodicity to be inherited by the embedded Calabi-Yau 3-fold, consider modifying~\eqref{e:dblPb} by ``splitting'' the $[m\pMod2]$-periodicity preserving degree ``2'' into:
\begin{equation}
  \Tw{Y}_{m,n}\in
  \K[{c||cc@{~~}c|c}
  {\IP^2 & 1 & 0 & 1 & \piC{\put(-2,1){\color{blue}\oval(12,5)}}1  \\
   \IP^1 & m & 0 & 1 & 1{-}m\\[1pt]
    \hdashline[1pt/1pt]\rule{0pt}{12pt}
   \IP^3 & 0 & 1 & 0 & \piC{\put(1,1){\color{blue}\circle{5}}}3  \\
   \IP^1 & 0 & n & 1 & 1{-}n\\}]_{18(m{-}6)}^{(4,58{-}9m)},\quad
  \Tw{Y}_{m,n}\subset \Tw{G}_{m,n}\in
  \K[{c||cc@{~~}c}
  {\IP^2 & 1 & 0 & 1 \\
   \IP^1 & m & 0 & 1 \\[1pt]
    \hdashline[1pt/1pt]\rule{0pt}{12pt}
   \IP^3 & 0 & 1 & 0 \\
   \IP^1 & 0 & n & 1 \\}].
 \label{e:nNo-m}
\end{equation}
The resulting $(m,n)$-sequence of Calabi-Yau 3-folds is however only simply, $[n\pMod3]$-periodic. The $m$-periodicity is broken by the lack of the first $\IP^2$-degree of ``2'' in the last constraint --- having split ``2''$\to$``$1\,|\,1$'' as highlighted by the oval; this also makes the Euler number and $h^{2,1}$ $m$-dependent. In fact, techniques detailed in Appendices~\ref{s:HFm} and~\ref{s:gTESS} permit computing $\dim H^0(\Tw{G}_{m,n},\cQ)=10(7{-}m)$. That is, $\Tw{G}_{m,n}$ has holomorphic anticanonical sections with which to define $\Tw{Y}_{m,n}$ only for $m\in[0,6]$, and the double sequence~\eqref{e:nNo-m} terminates in the $m$-direction while remaining infinite but $[n\pMod3]$-periodic in the $n$-direction. Of the so-defined $\Tw{Y}_{m,n}$, most (those with $2\leqslant m\leqslant6$) are gCICYs.

Finally, ``spoiling'' also the $n$-periodicity in the same way, we have the double sequence
\begin{equation}
  \breve{Y}_{m,n}\in
  \K[{c||cc@{~~}c|c}
  {\IP^2 & 1 & 0 & 1 & \piC{\put(-2,1){\color{blue}\oval(12,5)}}1  \\
   \IP^1 & m & 0 & 1 & 1{-}m\\[1pt]
    \hdashline[1pt/1pt]\rule{0pt}{12pt}
   \IP^3 & 0 & 1 & 1 & \piC{\put(-2,1){\color{blue}\oval(12,5)}}2  \\
   \IP^1 & 0 & n & 1 & 1{-}n\\}]_{-2 (40+(m-2)(n-3))}
                                ^{(4,44+(m-2)(n-3))},\quad
  \breve{Y}_{m,n}\subset \breve{G}_{m,n}\in
  \K[{c||cc@{~~}c}
  {\IP^2 & 1 & 0 & 1 \\
   \IP^1 & m & 0 & 1 \\[1pt]
    \hdashline[1pt/1pt]\rule{0pt}{12pt}
   \IP^3 & 0 & 1 & 1 \\
   \IP^1 & 0 & n & 1 \\}],\qquad
 \label{e:No-mn}
\end{equation}
which is periodic in neither $m$ nor $n$, and the Euler number depends on both $m$ and $n$. By direct computation as in Appendices~\ref{s:HFm} and~\ref{s:gTESS}, we find that the number of sections available for constructing the last constraint varies with both $m$ and $n$ and equals
\begin{equation}
   \dim H^0(\breve{G}_{m,n},\cQ)=3(15{-}3m{-}2n) \geqslant0\quad\text{for}\quad
   3m{+}2n\leqslant15,
\end{equation}
where $\cQ=\cO(1,1{-}m,2,1{-}n)=\cK^*_{\sss\breve{G}_{m,n}}$. Although this is an aperiodic $(m,n)$-sequence of configurations and terminates so $3m{+}2n\leqslant15$ (beyond which there are no $\breve{G}_{m,n}$-sections of $\cQ$ to define $\breve{Y}_{m,n}$), it contains 24 models, 18 of which are gCICYs.

\subsection{A higher codimension subtlety}
\label{s:cod2}
As a further illustration of the prescriptive power of the homological algebra encoded by the exact and spectral sequences --- and the formulae~\eqref{e:qm} and~\eqref{e:q=efp} in particular --- consider the following configuration:
\begin{equation}
 \begin{array}{lrll}
 \MR2*{$Z \in
         \K[{c||cc|c}{\IP^4&0&1&~~4\\\IP^2&2&2&-1\\}]^{(2,86)}_{-168}:\quad
  \left\{\rule[-4.5ex]{0pt}{4ex}\right.$}
  & s(y)&:= s_{(ij)}\,y^iy^j,\\[0mm]
  & p(x,y)&:= p_{(ij)}(x)\,y^iy^j &= p_{a\,(ij)}\,x^a\,y^iy^j,\\[1mm]
  & q(x,y)&:= \frac{q^i(x)}{y^i}  &= q^i_{(abcd)}\frac{x^ax^bx^cx^d}{y^i}.
 \end{array}
 \label{e:P24|22-1}
\end{equation}
The intermediate --- and regular --- complete intersection $M=\{s(y)=0=p(x,y)\}\subset \IP^4\,{\times}\,\IP^2$ has the standard Koszul resolution of its anticanonical bundle $\cQ=\cO(\mM{~\>4\\-1\\})$, and we list it with the associated spectral sequence:
\begin{equation}
\begin{array}{c|c@{\quad~}c@{\quad}c|c}
   & \cO\pM{~~3\\-5\\}
      \piC{\put(0,2){\vector(4,1){10}} \put(4,4.5){$\SSS p$}
           \put(0,1){\vector(4,-1){10}} \put(4,-3){$\SSS s$} }
   & \MM{ \cO\pM{~~4\\-3\\}\\[1mm]
          \cO\pM{~~3\\-3\\}\\ }
      \piC{\put(1,4){\vector(4,-1){10}} \put(6,4){$\SSS s$}
           \put(1,-2){\vector(4,1){10}}  \put(6,-3.5){$\SSS p$} }~~
   &\quad\str{13}\cO\pM{~~4\\-1\\}
   & \cQ_M \\[4mm]
 \hline\rule{0pt}{2.5ex}
 0. & 0 & 0 & 0 & H^0(M,\cQ) \\ 
 1. & 0 & 0 & 0 & H^1(M,\cQ) \\ 
 2. & \{\ve_{\9}^{ij(k}\vf^{l_1l_2)}_{(abc)}\}
     \piC{ \put(.5,2){\vector(2,1){5}}  \put(1.5,4.5){$\SSS p$}
           \put(.5,1){\vector(2,-1){5}} \put(1.5,-3){$\SSS s$}
           \color{blue}\mdot(-7,4){14}{45}
                       \mdot(37,15){0}{8}}
       & \MM{ \{\ve_{\9}^{ijk}\f_{(abcd)}\} \\[1mm]
              \{\ve_{\9}^{ijk}\f_{(abc)}\} \\ }
     \piC{\color{blue}\mdot(0,5){21}{12}
                      \mdot(-1,-1){40}{16}
                       \mdot(11.5,9.5){0}{8}}
           & 0 & 0 \\ 
 3. & 0 & 0 & 0 & 0 \\[-1mm]
\vdots\> & \vdots & \vdots & \vdots & \vdots \\ 
 \hline
 \MC5l{\text{\footnotesize $H^1(M,\cQ)=0$ for sufficiently generic $s(y)$ and $p(x,y)$.}}
\end{array}
 \label{e:24|-14}
\end{equation}
If $s_{(ij)}$ and $p_{a\,(ij)}$ in~\eqref{e:P24|22-1} are chosen sufficiently generically and the combined $p\,{+}\,s$-mapping is surjective, nothing is mapped to $H^1(M,\cQ)$. The kernel of this mapping (domain elements that are annihilated) is then 105-dimensional and is identified with $H^0(M,\cQ)$, which is thus parametrized by the tensor coefficients $\vf^{(ij)}_{(abc)}$ satisfying the kernel constraints:
\begin{equation}
 \big\{\vf^{(jk)}_{(abc)}:~~
        \vf^{(jk)}_{(abc)}s^{\9}_{(jk)}=0=f^{(jk)}_{(abc}p^{\9}_{d)(jk)}\big\}
\end{equation}
These coefficients are then used to define the degree-$(\mM{~\>4\\-1\\})$ rational sections by means of a double contraction
\begin{equation}
  \g^{\,i\9}_{(abcd)}
  = p_{(j_1j_2)(a}\>\ve^{ij_1k_1\9}_{\9}\vf^{(j_2k_2)}_{bcd)}\,s^{\9}_{(k_1k_2)}.
 \label{e:q=sefp}
\end{equation}
The appearance of the $\ve^{ijk}$-symbol again introduces relative signs, much as $\ve^{ij}$ does in~\eqref{e:4qxy}.

A quick comparison of this prescription with~\eqref{e:q=efp}, and a corresponding comparison of the chart~\eqref{e:24|-14} with~\eqref{e:mH*Q} reveals the following:
\begin{enumerate}\itemsep=-3pt

 \item The $\ve$-symbols in~\eqref{e:q=efp} and~\eqref{e:q=sefp} are dual to the volume-forms of $\IP^1$ and $\IP^2$, respectively, owing to the fact that these sections stem, respectively, from 1- and 2-forms in~\eqref{e:mH*Q} and~\eqref{e:24|-14}, respectively.
 
 \item The representatives~\eqref{e:q=efp} are obtained by a single contraction with $p_{a\,(ijk_1\cdots k_n)}$---as dictated by the resolution map indicated in the ``header'' of the chart~\eqref{e:mH*Q} and since $F_m$ is a codimension-1 hypersurface in $\IP^4\,{\times}\,\IP^1$. By contrast, the representatives~\eqref{e:q=sefp} include a double ``pullback'' contraction, using each of the two tensor coefficients $s_{(ij)}$ and $p_{a\,(ij)}$. Again, this is as dictated by the sequence of resolution maps indicated in the ``header'' row of the chart~\eqref{e:24|-14} and owing to $M$ being a codimension-2 intersection $\{s(x,y)=0=p(x,y)\}$ of hypersurfaces in $\IP^4\,{\times}\,\IP^2$.
 
\end{enumerate}

Owing to Corollary~2 of Ref.~\cite{rGHL-Hog} (see also Ref.~\cite[Lemma~2.1, p.\,54]{rBeast}), we have the sequence of relation
\begin{equation}
  \K[{c||cc|c}{\IP^4&0&1\,&4\\\IP^2&2&n&1{-}n\\}]
  ~\cong~
  \K[{c||c|c}{\IP^4&1\,&4\\\IP^1&2n&2{-}2n\\}],
 \label{e:1>2}
\end{equation}
so that the configuration in the left-hand side $n$-sequence correspond to even-$m$ configurations in our main example sequence~\eqref{e:Seq}. This relationship derives from the fact that a generic quadric in $\IP^2$ is isomorphic to $\IP^1$. However, one must be careful in using this relation since it does not include an isomorphism of the integral cohomology. The straightforward Chern class computation for the quadric $[\IP^2||2]$ and the hyperplane in $[\IP^2||1]$ which equals $\IP^1$,
\begin{equation}
   \big(c[\IP^2||2] = 1+J_3\big) ~\cong~
   \big( c[\IP^1] = 1+2J_2 \big) ~\approx~
   \big(c[\IP^2||1] = 1+2J_4\big),
\end{equation}
implies that $(J_2\,{=}\,J_4)=\frac12J_3$. That is, the generator of $H^2\big([\IP^2||2],\ZZ)$ is identified with the square of the generator of
 $H^2\big([\IP^2||1],\ZZ)\approx H^2(\IP^1,\ZZ)$.
 Indeed, using the corresponding {\em\/fractional\/} generator, we calculate the intersection numbers~\cite[p.\,178]{rBeast}:
\begin{equation}
  \ssK[{c||cc|c||c@{~~}r@{~~}r}{\IP^4&0&1&4&1&1&1\\\IP^1&2&n&1{-}n&0&0&0}]=2+6n
   \quad\text{and}\quad
  \ssK[{c||cc|c||c@{~~}r@{~~}r}{\IP^4&0&1&4&1&1&0\\
                    \IP^1&2&n&1{-}n&0&0&
                     \piC{\put(1,.75){\color{blue}\circle{5}}}{\ttt\frac12}}]=4.
 \label{e:1/2k3}
\end{equation}
Using the standard result $c_2=\big(6J_1^{~2}+(4{-}3n)J_1J_3+(1{-}n{+}n^2)J_3^{~2}\big)$, we have
\begin{alignat}9
 [(aJ_1{+}b\,\piC{\put(3,1){\color{blue}\circle{7}}}{\ttt\frac12}J_3)^3]
  &= 2a^3+3a^2(\2{4b{+}2na}), \label{e:1/2J3}\\[2mm]
 C_2[(aJ_1{+}b\,\piC{\put(3,1){\color{blue}\circle{7}}}{\ttt\frac12}J_3)]
  &= 44a +6(\2{4b{+}2na}). \label{e:1/2c2J}
\end{alignat}
Comparing~\eqref{e:1/2J3} and~\eqref{e:1/2c2J} respectively with~\eqref{e:k3Xm} and~\eqref{e:p1Xm} proves that $m=2n$, and the left-hand side sequence of configurations in~\eqref{e:1>2} indeed captures only the even-$m$ configurations in the original series~\eqref{e:Seq}. More generally, configuration relations of the form
\begin{equation}
   X\in\K[{c||c;{1.5pt/1pt}c}{\mathbb{X}&0&\mathbb{A}\\
                              \hdashline[1.5pt/1pt]\IP_y^2&2&\mathbb{B}\\}]
   ~\cong~
   \K[{c||c}{\mathbb{X}&\mathbb{A}\\\hdashline[1.5pt/1pt]
              \IP_z^1&2\mathbb{B}\\}]\ni \mathring{X}
   \qquad\Rightarrow\qquad
   {\ttt\frac12}J_y =J_z
\end{equation}
imply a homeomorphism but not a diffeomorphism between $X$ and $\mathring{X}$; in particular, the classical cohomology rings of $X$ and $\mathring{X}$ agree only upon a (non-integral) rational basis change.

\providecommand{\href}[2]{#2}\begingroup\raggedright\endgroup

\end{document}